\documentclass[twocolumn]{aastex6}

\usepackage[T1]{fontenc}
\usepackage{ae,aecompl}

\usepackage{graphicx}	
\usepackage{amsmath}
\makeatletter
\def\env@matrix{\hskip -\arraycolsep 
  \let\@ifnextchar\new@ifnextchar
  \array{*{\c@MaxMatrixCols}c}}
\makeatother

\usepackage{amssymb}	
\usepackage{csquotes}

\begin{document}
\title{AGN Variability Analysis Handbook}
\author{
	Jackeline Moreno$^{1,2}$,
	Michael S. Vogeley$^{1}$,
	Gordon T. Richards$^{1}$
	\\
	$^{1}$Drexel University, 3141 Chestnut Ave, Philadelphia, Pennsylvania 19104, USA \\
    $^{2}$ LSSTC Data Science Fellowship Program Fellow}

\date{Submitted; Received 2018 Oct 30}

	\begin{abstract}
This work develops application techniques for stochastic modelling of Active Galactic Nuclei (AGN) variability as a probe of accretion disk physics.  Stochastic models, specifically Continuous Auto-Regressive Moving Average (CARMA) models, characterize lightcurves by estimating delay timescales that describe movements away from and toward equilibrium (mean flux) as well as an amplitude and frequency of intrinsic perturbations to the AGN flux. We begin this tutorial by reviewing discrete auto-regressive (AR) and moving-average (MA) processes, we bridge these components to their continuous analogs, and lastly we investigate the significance of timescales from direct stochastic modelling of a lightcurve projected in power spectrum (PSD) and structure function (SF) space.  We determine that higher order CARMA models, for example the Damped Harmonic Oscillator (DHO or CARMA(2,1)) are more sensitive to deviations from a single-slope power-law description of AGN variability; unlike Damped Random Walks (DRW or CAR(1)) where the PSD slope is fixed, the DHO slope is not. Higher complexity stochastic models than the DRW capture additional covariance in data and output additional characteristic timescales that probe the driving mechanisms of variability. 

	\end{abstract}

	\maketitle

	
	
	\section{Introduction}

This work reviews the traditional statistical toolkit for time series analysis of active galactic nuclei (AGN) variability with an emphasis on stochastic modelling.  We aim to provide a practical understanding of simple stochastic models such as damped random walks (DRW) and (perturbation-driven)
damped harmonic oscillators (DHO).  Stochastic models are applied widely to economic and engineering data, but their interaction with gappy astronomical data is still in its relative infancy.    
The goal of this work is to provide an accessible description of a family of stochastic models known as auto-regressive moving-average (ARMA) models, specifically for the benefit of early career researchers or more experienced researchers interested in learning to apply these methods to astrophysical data.  Included is discussion of ARMA models both in their discrete form and (more importantly to astronomers) in their continuous form (CARMA).   We discuss why CARMA models of higher order than the DRW (such as the DHO) may be useful and appropriate, and also when the data do not justify such models.  For more details on the topics discussed herein we refer the reader to \citet{panditWu1975},  \citet{GrangerMorris}, \citet{mills2012}, \cite{BoxJenkins2015}, \citet{Kelly14}, \citet{Vish2017}, \citet{Feigelson}.

Early science investigations of AGN variability employed simple stochastic models 
to characterize changes in X-ray brightness (milli-second to month timescales) (\citealt{UttleyPSD}, \citealt{Vaughan2003}, \citealt{Uttley2005}, \citealt{McHardy2006Natur}). A benefit of investigating variability in the optical is to expand the population of AGN beyond small-areas or limited to X-ray bright sources (which tend to be biased towards relatively low-luminosity and low-redshift). 
A broader goal of the community is to understand if variability is casually connected from the inner X-ray corona to the outer UV/optical/infrared/radio-emitting annuli of the disk (\citealt{Pringle1981}, \citealt{Done2018}).

Optical variability of AGN is observed in the range of roughly 3--50\% with the median value on the order of $20\%$ (\citealt{Sesar2007}, \citealt{MacLeod2008}). 
Interest in expanding the statistical toolkit to read information encoded in a stochastic lightcurve is motivated by a desire to learn about the physical mechanisms generating energy over a large range of observed timescales. Measures of variability characteristics potentially relate to the diversity of Spectral Energy Distributions (SED) of AGN and may improve our understanding of black hole mass, fueling, and spin in addition to accretion disk geometries (\citealt{Krawczyk2013}, \citealt{Richards2006}, \citealt{Brandt2018}, \citealt{Sartori2018}). 

This paper introduces ARMA models as building blocks with an emphasis on visualizing the reconstruction of lightcurve shapes (Sections 2-4). We build a bridge from discrete to continuous ARMA models to facilitate interpretation of the continuous model parameter space where subregions reveal different dynamics (or shapes) of lightcurves (Section 5).  Section 6 and 7 delve into deeper explanations of CARMA differential equations necessary for navigating the parameter space to conduct an ensemble study of AGN variability.  The remaining sections review more traditional analyses including autocorrelation, structure functions, spectral analysis, and impulse-response functions---all in relation to CARMA models. Specifically, we emphasize the connections between these methods and offer pitfall discussions based on modelling real astrophysical data (Moreno et al. 2019 in prep).   This handbook is meant to advance the application of stochastic models to astrophysical data and to contextualize such methods with more traditional analyses used to study AGN and stars in the time domain. 

\section{Auto Regression}    
An auto-regressive (AR) process describes a system whose future value can be predicted given a measure of its current state.  That is, AR is a form of forecasting. In a typical linear regression $(y=mx+b)$ we predict the value of a dependent variable $(y)$ based on the value of the independent variable $(x)$.  In an {\em auto-regressive} process we instead predict the future value of the system based on the past value (or values) of the system itself.  Thus the dependent and independent variables are the same, for example: 
\begin{equation}
x_{i} =\mu + \phi_{1}x_{i-1} + \epsilon_{i},
\label{eq:RW}
\end{equation}
where $\phi_1$ is the auto-regressive or \enquote{lag} coefficient that indicates how closely tied future values are to past values and where $\epsilon_i$ represents a source of noise. 

\begin{figure}
    \plotone{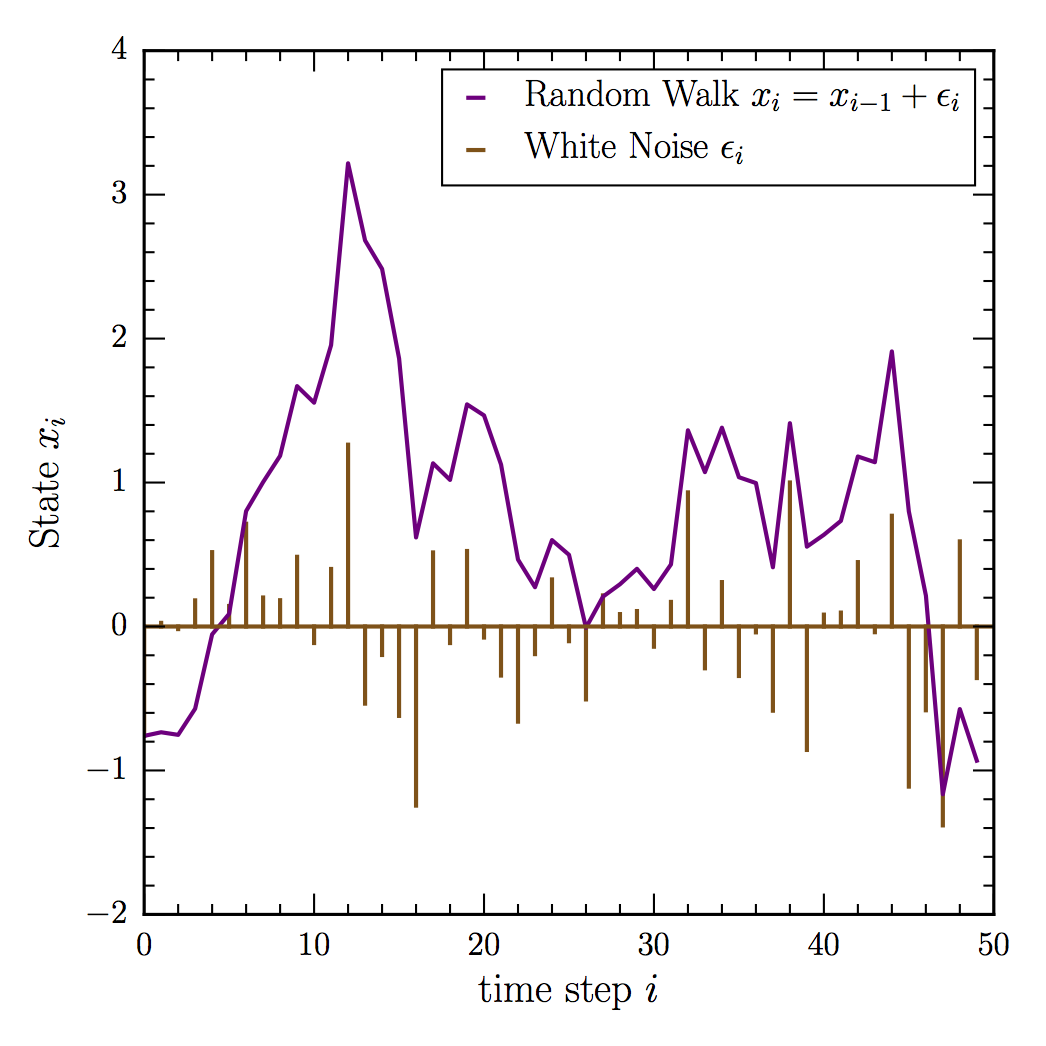}
   	\caption{Illustration of a random walk in the context of an AR(1) model.  White noise disturbances (represented by the gold lines) are shown as a train of shocks that randomly change the state of the system (solid purple).  Random walks driven by white noise are the simplest stochastic model and are a special case of AR(1) models.  Note that the amplitude of the system can grow arbitrarily large given enough time.}
	\label{fig:RW}
\end{figure}

Equation~\ref{eq:RW} is a first-order or \enquote{AR(1)} model where the current state of the system is linearly dependent on the immediately prior state plus a random perturbation $\epsilon_i$ (conventionally drawn from a Gaussian distribution). These random perturbations are not external to the system like measurement error;
$\epsilon_i$ represents a true perturbation mechanism intrinsic to the system.  A special case of an AR(1) process, known as a \enquote{random walk}, has $\phi_1=1$ with $\epsilon_i$ represented by \enquote{white noise} (see \S~\ref{sec:PSD}).  A random walk process is illustrated in Figure~\ref{fig:RW} where we show both the white noise impulses (the ``shock train") and the resulting state of the system, $x_i$, at each time step.  

Since the current state of the system depends on the previous state, it is possible for $x_i$ to grow beyond the amplitude of the shocks themselves if there are a number of consecutive shocks with the same sign.  This is true not just for the random walk shown in Figure~\ref{fig:RW}, but generically for an AR(1) process with $\phi_1>1$ (or for $\phi_1<-1$).
Note that the {\em change} in a random walk is equivalent to white noise: $x_i - x_{i-1} = \epsilon_i$.

\begin{figure*}
\plottwo{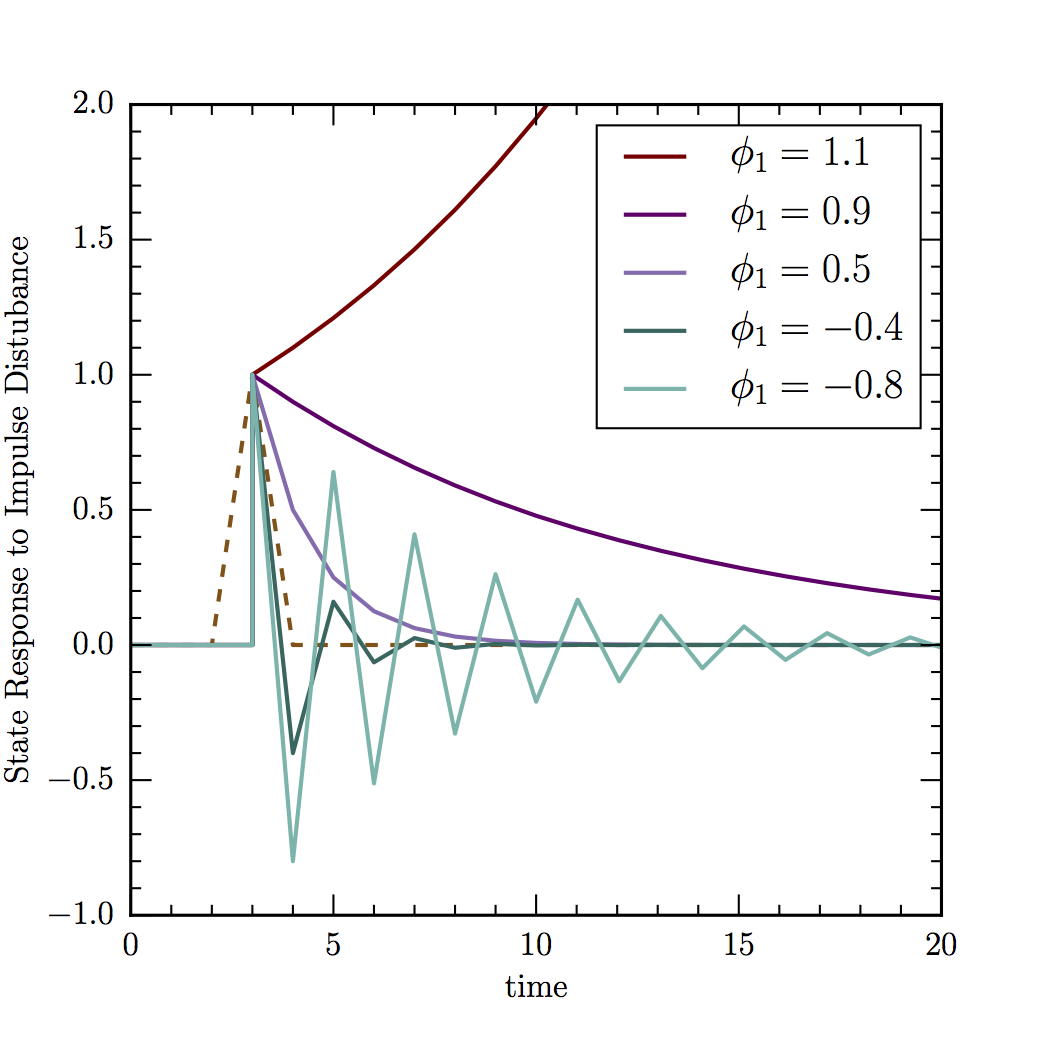}{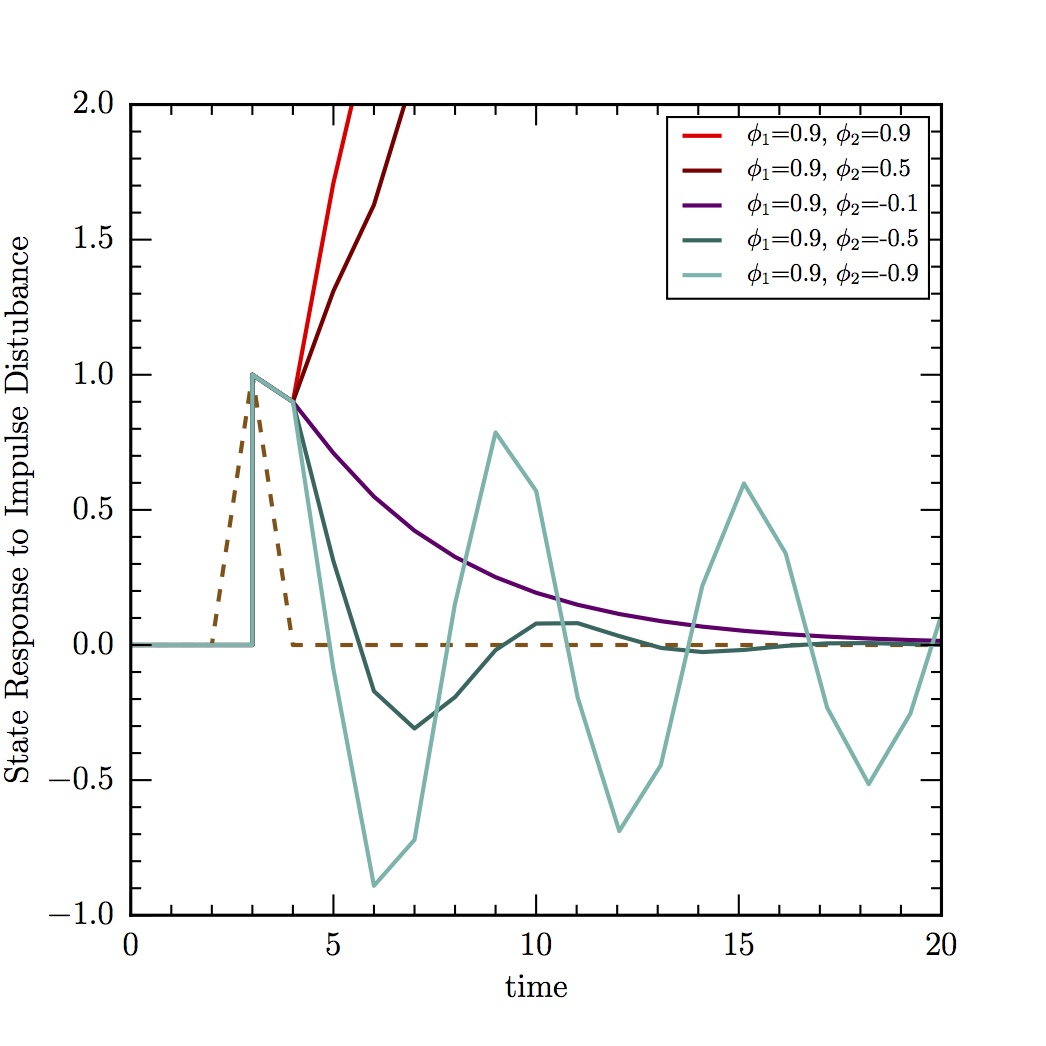}
   	\caption{Response of an AR system to a single impulse (gold) for a range of lag coefficients, $\phi_i$.  (Left:) An AR(1) with a single unit impulse (shown as triangle for visibility) at time step 4.  $\phi_1>1$ leads to a runaway, while $-1<\phi_1<1$ reverts to the mean, the speed of which depends on the magnitude of $\phi_1$.  (Right:) As with the left panel except for a second-order, AR(2), process. If both $\phi_i$ values are positive, then the system runs away (no matter how small the coefficients).  $\phi_1 >0$ is permitted if $\phi_2<0$, resulting in decaying oscillatory motion.}
	\label{fig:AR}
\end{figure*}

The state of a generic AR system, $x_{i}$, is easier to understand by visualizing the evolution forward in time after a single shock.  
The left panel of Figure~\ref{fig:AR} presents examples of a first-order AR processes for a range of coefficients, $\phi_1$ (Equation~\ref{eq:RW}).
The lag coefficient simply interpolates from one state (data point) to the next by acting as the slope between adjacent time step pairs.
First order AR models have dynamics that encompass 
three domains:
(1) $\phi_1 < 0$ results in a decaying sawtooth, (2) $0<\phi_1<1$ results in fast to gradual decay, and (3) $\phi_1 >1$ leads to unbounded growth. 

The general notation for a $p_{th}$-order AR process is 
\begin{equation}
x_{i} = \mu + \sum^{p}_{1}{\phi_{p}x_{i-p}} + \epsilon_{i}.
\label{eq:AR}
\end{equation}
For example, in an AR($p=2$) process, the state of the system depends on two lag terms (and thus has two timescales):
\begin{equation}
x_{i} =\mu +\phi_{1}x_{i-1}+ \phi_{2}x_{i-2}+ \epsilon_{i}.
\label{eq:AR2}
\end{equation}
The response to a single impulse in such a system is shown in the right panel of Figure~\ref{fig:AR}. In this example, the first lag coefficient is held constant, while the second lag term is varied.  We highlight three types of dynamics in parallel with the first order models in the left panel of Fig. 2. When $\phi_2 \rightarrow 0$ (and $\phi_1$ necessarily becomes relatively more significant than $\phi_2$) the lag response reverts to the purple, gradually decaying curves of the AR(1) process on the left panel.  Negative values of the second lag coefficient, $\phi_2 < 0 $, result in decaying oscillatory motion.  Finally, positive values of the second lag coefficient, $\phi_2 > 0$, lead again to unbounded growth due to impulse perturbations.

In AR processes, the lag coefficients provide short-memory-dependent structure in the form of positive or negative correlations to the previous state. Increasing the order of an AR process is equivalent to increasing the memory of past events in the system, akin to inertia.  To achieve a long-term dependence requires expanding the sum to more and more AR terms. Long-term dependence can also be recognized as a drift away from the mean, implying non-constant (i.e. time-dependent) coefficients referred to as ``non-stationarity". A stationary system in this context is a system for which the mean does not drift with time. We will return to stationarity in Section~\ref{sec:ACF}.

In the context of this paper, it is important to realize that the full domain of AR coefficients is not relevant for AGNs.   Appropriate representations of AGN systems exhibit stability such that they return to the mean and have continuous analogs, which we will discuss in Section~\ref{sec:bridge}. 
One of the goals of this paper is to achieve a better understanding of the transition from gradual decay to decaying oscillatory behavior within the AR domain and how it may be used as a probe of AGN variability.

\section{Moving Averages}
\begin{figure} 
    \plotone{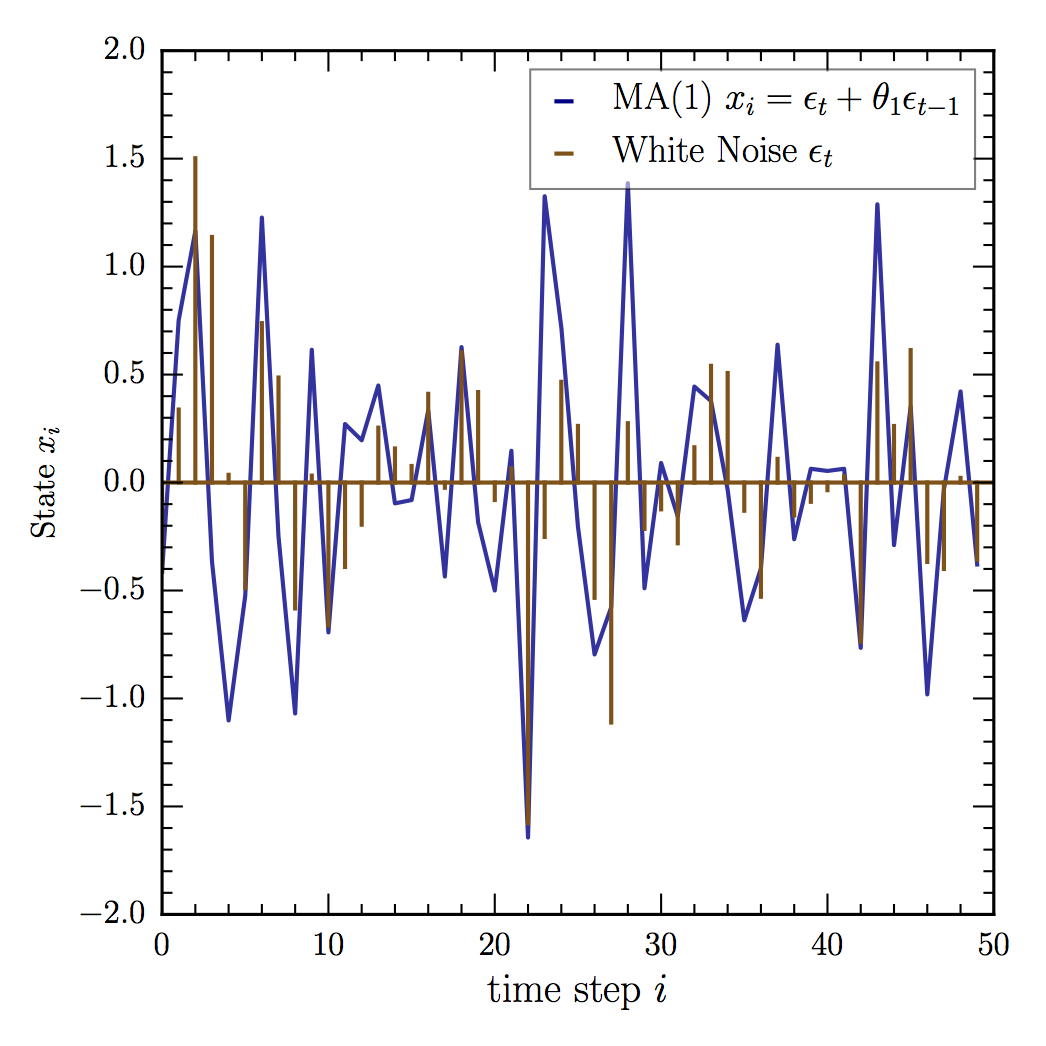}
   	\caption{A moving average process. The same train of white noise impulses (gold) used to generate the random walk in Figure~\ref{fig:RW} and a resulting moving average ($\theta_1=-1$; blue line) process are over-plotted. The MA coefficient determines the amplitude and delay between perturbations. The MA process is bounded by the amplitude of the individual shocks (unlike AR processes).  }
	\label{fig:MA}
\end{figure}

In a moving average (MA) process, instead of the future value depending on the previous value(s) of the system, the future state dependS only on previous {\em shocks} to the system,  $\epsilon_i$, so that
\begin{equation}
x_i= \epsilon_i+\theta_1\epsilon_{i-1}.
\label{eq:MA}
\end{equation}
In the simplest MA process (order 1 with $\theta_1 \neq0$), the system depends not only on the current shock, but also the previous one.  Figure~\ref{fig:MA} presents the state of an MA system driven by white noise with $\theta_1 = -1$.   Unlike Figure~\ref{fig:RW}, which has the same underlying shock train, the MA system in Figure~\ref{fig:MA} is bounded by the size of the individual shocks.  

\begin{figure*} 
    \plotone{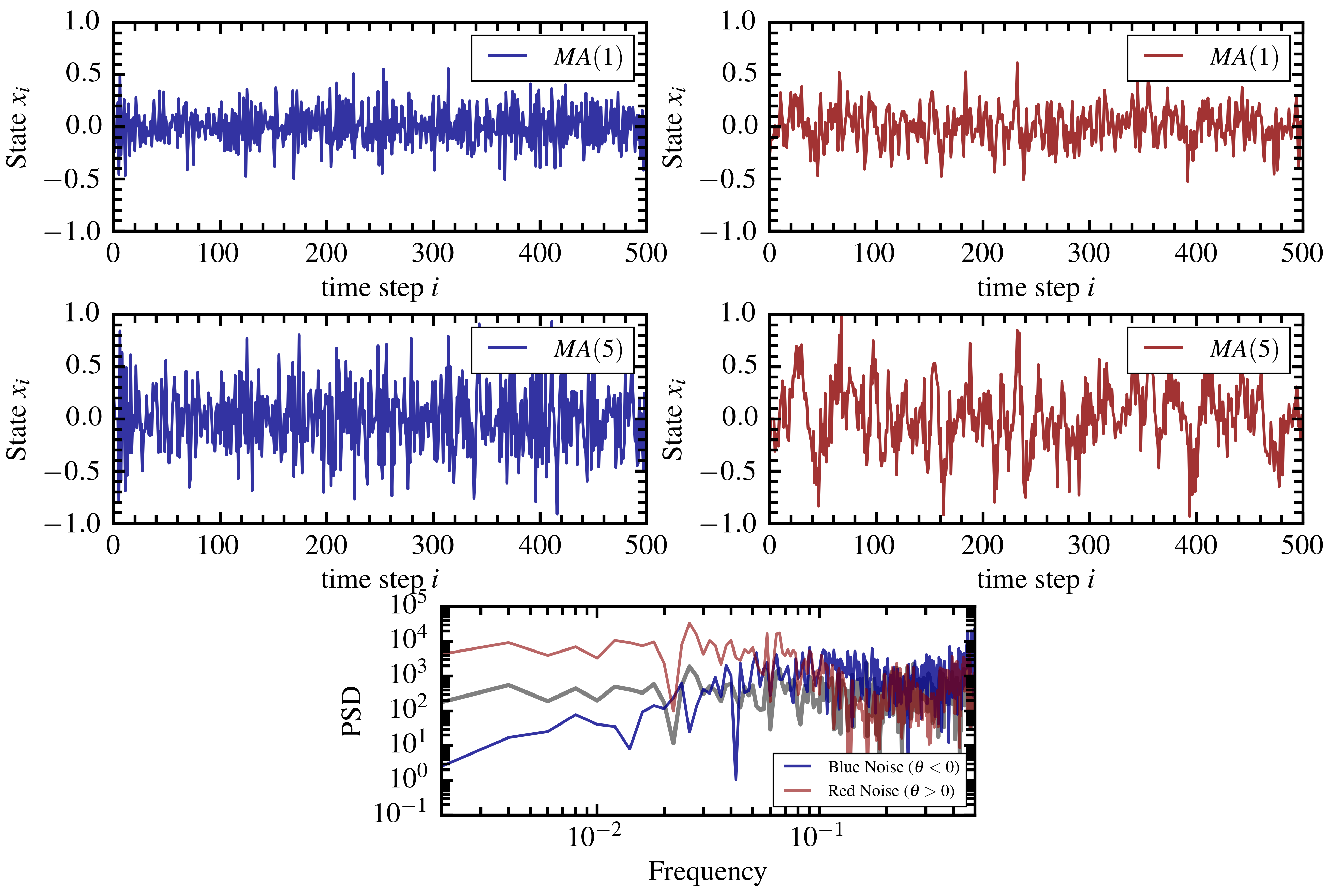}
   	\caption{Moving Average models of different orders. Left: MA processes with all negative coefficients, $\theta_i <0$.  Right: MA processes with the same amplitude as in the left panels, but with all positive MA coefficients, $\theta_i >0$.  All MA processes are generated with the same (white noise) shock train. Bottom: Power spectral density (PSD; see \S~\ref{sec:PSD}) of the input white noise spectrum (grey line) and for $\theta_i <0$ (blue) and $\theta_i > 0$ (red).  MA processes with $\theta_i<0$ exhibit a ``blue" noise spectrum (more power at higher frequencies), while for $\theta_i>0$ power is transferred to low frequencies resulting in a ``red" power spectrum. 
   	The number and magnitude of MA coefficients in a model determines the amplitude and the frequency of disturbances.}
	\label{fig:MAmany}
\end{figure*}

Since the state of an MA process depends only on previous shocks and not previous values of the system, the effects of a shock are more limited in time.  Thus the equivalent of Figure~\ref{fig:AR} for an MA process would be rather subtle and not be very interesting.  Instead, in Figure~\ref{fig:MAmany} we show what happens to the system as a result of increasing the order (for both positive and negative values of $\theta$).  As with AR processes, higher order generic MA($q$) processes are given by 
\begin{equation}
x_{i} = \sum^{q}_{1}{\theta_{q}\epsilon_{i-q}} + \epsilon_{i}.
\end{equation}
The effect of coefficients with $\theta_q < 0$ and $\theta_q > 0$ are shown in the left and right panels of Figure~\ref{fig:MAmany}, respectively. MA models are intrinsically shorter-term than AR but they also characterize correlation structure of a different flavor. 

When $\theta_q < 0$ each shock is included in the next time step with the opposite sign, making the model highly erratic (high frequency).  On the other hand, when $\theta_q > 0$ the result of each shock is more persistent---even more so when the order is increased as can be seen going from the upper to the middle panel on the right-hand side of Figure~\ref{fig:MAmany}. The number and magnitude of MA terms in the model determines the amplitude and delay between perturbations.  Thus, the MA coefficients determine whether the process $x_{i}$ is highly erratic or rather quiet by modulating the delay between perturbations, so MA coefficients are sometimes said to act as noise filters. 

While white noise exhibits an equal distribution of power across all frequencies, MA {\em reprocessing} of white noise input can transfer power to higher (or lower) frequencies---despite the input and output processes having equal means and standard deviations. 
The standard deviation cannot distinguish whether higher amplitude deviations happen at low frequency or high frequency. This information is exactly what the MA process describes and it can be leveraged to help reconstruct a lightcurve shape more accurately. 

MA coefficients encode frequency domain information that is traditionally obtained from a power spectrum.   Thus, the differences between negative and positive values of $\theta$ are better illustrated in frequency domain (in terms of correlated or ``colored" noise) than in the time domain as shown in the bottom panel of Figure~\ref{fig:MAmany}. The slope and steepness of the power spectral density (PSD) of a stochastic process manifests as the ``smoothness" of a noisy process. When $\theta_1 < 0$, large amplitude perturbations are closely concentrated together in time (at high frequency).  When $\theta_1 > 0$ the opposite effect results in high amplitude perturbations spread out across lower frequencies.  MA processes with positive and negative valued $\theta_i$ are contrasted in the top two rows of Figure ~\ref{fig:MAmany}. The differences are more apparent in the frequency domain, however it it useful to recognize the signature of this type of correlation in the time domain. 
In Section~\ref{sec:PSD}, we will find that, in the context of AGN variability, we are highly interested in applications of MA processes that produce excess power at higher frequencies, i.e. ``blue" noise as shown in the left panels of Figure~\ref{fig:MAmany}.

While MA processes are intrinsically short-term, we note that it is possible to rewrite a first order auto-regressive model as a moving average process with infinite terms: AR(1) = MA($\infty$) with all positive $\theta_i$ (Box and Jenkins 1970).  However, in practice, an MA model of a finite number of terms cannot exactly replicate an AR process. Together AR and MA processes can be thought of as building blocks of more complex processes.  In the next section, we summarize mixed processes called auto-regressive moving-average or ARMA processes (\citealt{BoxJenkins2015}).   Understanding the dynamics that these different building blocks describe will help with model application to astrophysical data and interpretation of coefficients.

\section{Mixed Models: ARMA}
\label{sec:mixed}
In this section, we explore how AR and MA components in discrete ``mixed" ARMA models can be used to characterize the shape and statistical complexity of a stochastic time series---such as an AGN light curve. ARMA models 
combine short-memory AR responses and MA inputs that govern the \enquote{amplitude} of random perturbations at different timescales.  Together AR and MA parameters reconstruct correlation structure and the degree of smoothness in noisy processes. In discrete ARMA models each index in the series evolves the process forward by a constant time step.  In continuous models (next section), the relative significance between coefficients interpolates behavior in sparse and gappy time-series data.  Familiarity with discrete models provides an intuition for the complexity of continuous models which are more practical for ``survey-quality" data like the Large Synoptic Survey Telescope (LSST; \citealt{Ivezic2008}) will provide.

Both AR and MA models can be used separately (or together) to describe stochastic processes.  For example AR is a good description of day-to-day stock prices while MA is a good description of intraday stock prices.
However, both in the case of stock prices and the AGNs in our analysis, when combining the models (ARMA analysis) it is useful to think of the MA model as the {\em input} to the AR model.  

\begin{figure*}
	\plottwo{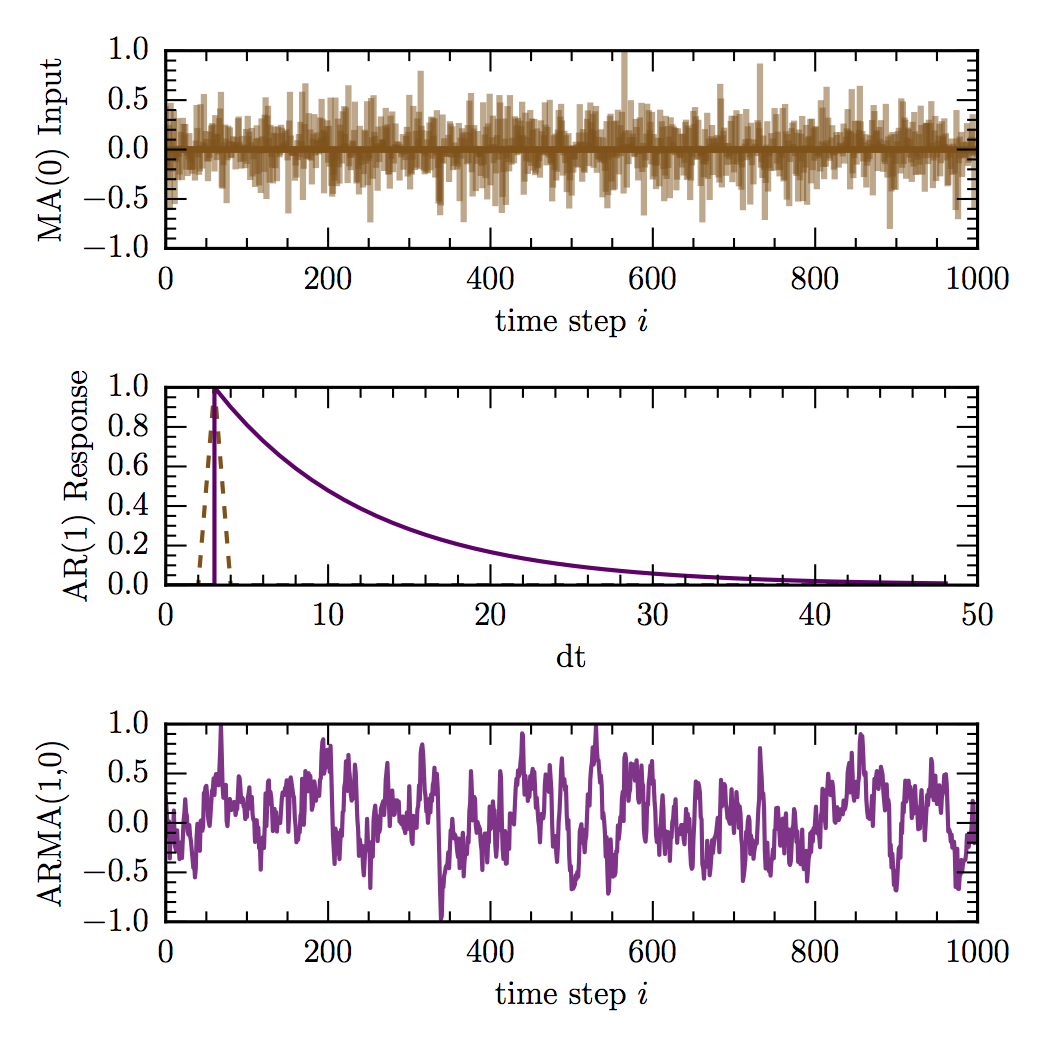}{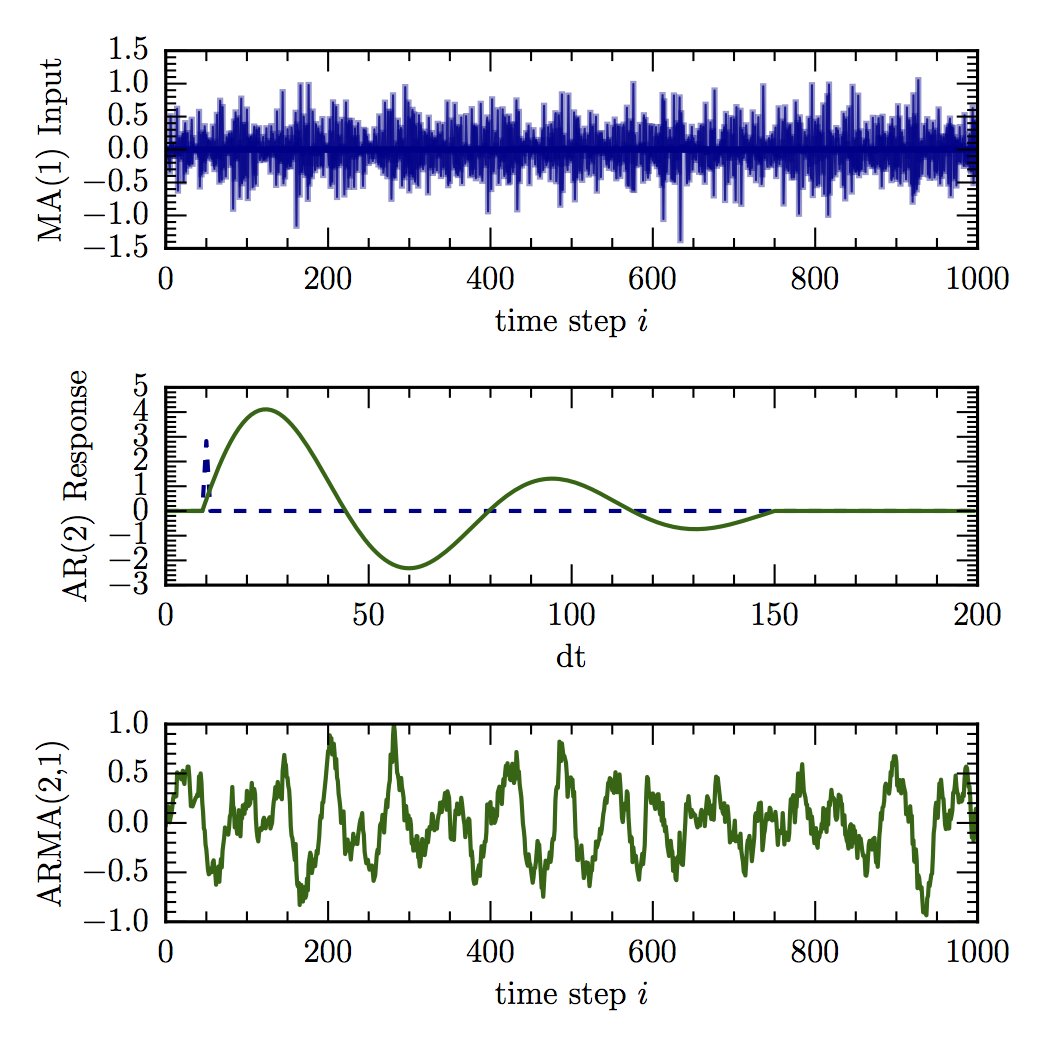}\\
   	\caption{Building blocks of ARMA(1,0) (left) and ARMA(2,1) (right) models.   A white noise input (top left) and an MA(1) input (derived from the same white noise; top right), the response of lag coefficient(s) (middle), and the final stochastic process (bottom panel). Input perturbations to the system are described by MA coefficients (or lack thereof) and AR lag coefficients describe how the system responds and dissipates energy back to the mean state.  Despite having the same input white noise, the final mixed processes (left vs.\ right) have noticeably different characteristics.}
	\label{fig:MixedModels}
\end{figure*}

Thus an AR(1) model describes a system with Brownian motion (random walk) character and a decay channel responsible for dissipating energy, while an MA process describes the energy {\em inputs}.   The convolution of these two components is an ARMA(1,0) process (Figure~\ref{fig:MixedModels}, left), where the $\epsilon_i$ term is now represented as an MA(0) process.

The next order mixed model arises from the convolution of an MA(1) process kicking the system and a second order AR response channel, ARMA(2,1),  (Figure~\ref{fig:MixedModels}, right), expressed as 
\begin{equation}
x_i = \mu + \phi_1 x_{i-1} + \phi_2 x_{i-2} + \theta \epsilon_{i-1} + \epsilon_i.
\end{equation}
With only a single lag term (p=1, thus $\phi_2=0$), an AR(1) model can only characterize the dissipation rate of impulse perturbations back towards the mean state.  With two lag terms, an AR(2) model can additionally characterize delays as the process moves \emph{away} from the mean state.  The green curve in the bottom panel of Figure 5 (right) exhibits more correlated structure in the form of asymmetric peaks and troughs. Thus, input and response components are the building blocks of more complex stochastic processes. 

If considering higher-order models for astrophysical applications, it is important to realize that $q < p$ is a condition of stationarity for continuous processes. When we explore continuous models (next section), we will relate ARMA parameters to characteristic timescales of variability in the continuous space.

\section{ARMA to CARMA Bridge}
\label{sec:bridge}

\subsection{Going From Discrete to Continuous}
Although the AGN light curves that we wish to study may be driven by (intrinsically) continuous processes, we could in principle perform our analysis using discrete ARMA processes {\em if} the light curves were sampled uniformly.  However, in practice, light curves are rarely sampled uniformly (due to seasonal and nightly gaps) and, even when they are, bad data points can disturb an otherwise uniform cadence.  In the case of LSST data, we expect that light curves will have a very non-uniform cadence.  Thus, we must treat our systems with the continuous mathematical analogs of discrete ARMA processes.  

To gain insight into the transition from discrete to continuous analysis, we remind the reader of the derivation of \enquote{$e$}, which figures prominently in this transition.   For example, consider the case of compound interest.  If we have a dollar and earn 1\% interest per year (r=0.01), we can write an AR(1) equation to describe the process as $$D_t = \phi_1 D_{t-1},$$
with  $$\phi_1 = 1+r .$$ 
If our interest payments are made $n$ times a year instead of at the end of the year, then we have $$\phi_1 = 1 + \frac{r}{n},$$
and our final account balance will be 
 $$D_{t=1 yr}= \phi_{1}^n D_{0}.$$
 If our payments are made {\em continuously} rather than in discrete intervals, we then have 
 $$ \lim_{n \to \infty} \phi^n = \lim_{n \to \infty} (1+\frac{r}{n})^n = e^r,$$   
so that after $t$ years, our balance is 
$$D_t = e^{rt}D_0.$$
Thus we can see how a discrete AR(1) process can morph into a continuous process: in the limit where the time interval gets very small compared to the observation length, the process approaches continuous behavior.
For $r>0$ we have exponential growth, just as we had with $\phi_1>1$ for a discrete AR(1) process.    If $r<0$ (compounded bank fees), then we have exponential decay.  Continuous-AR(1) models with $r<0$ are better known as damped random walks (DRW) and are commonly used models of AGN variability (\citealt{Kelly2009}, \citealt{KozlowskiDRW}, \citealt{MacLeod2011}, \citealt{MGwavelet}).

\subsection{The Stability Triangle}

\begin{figure}[!] 
	\plotone{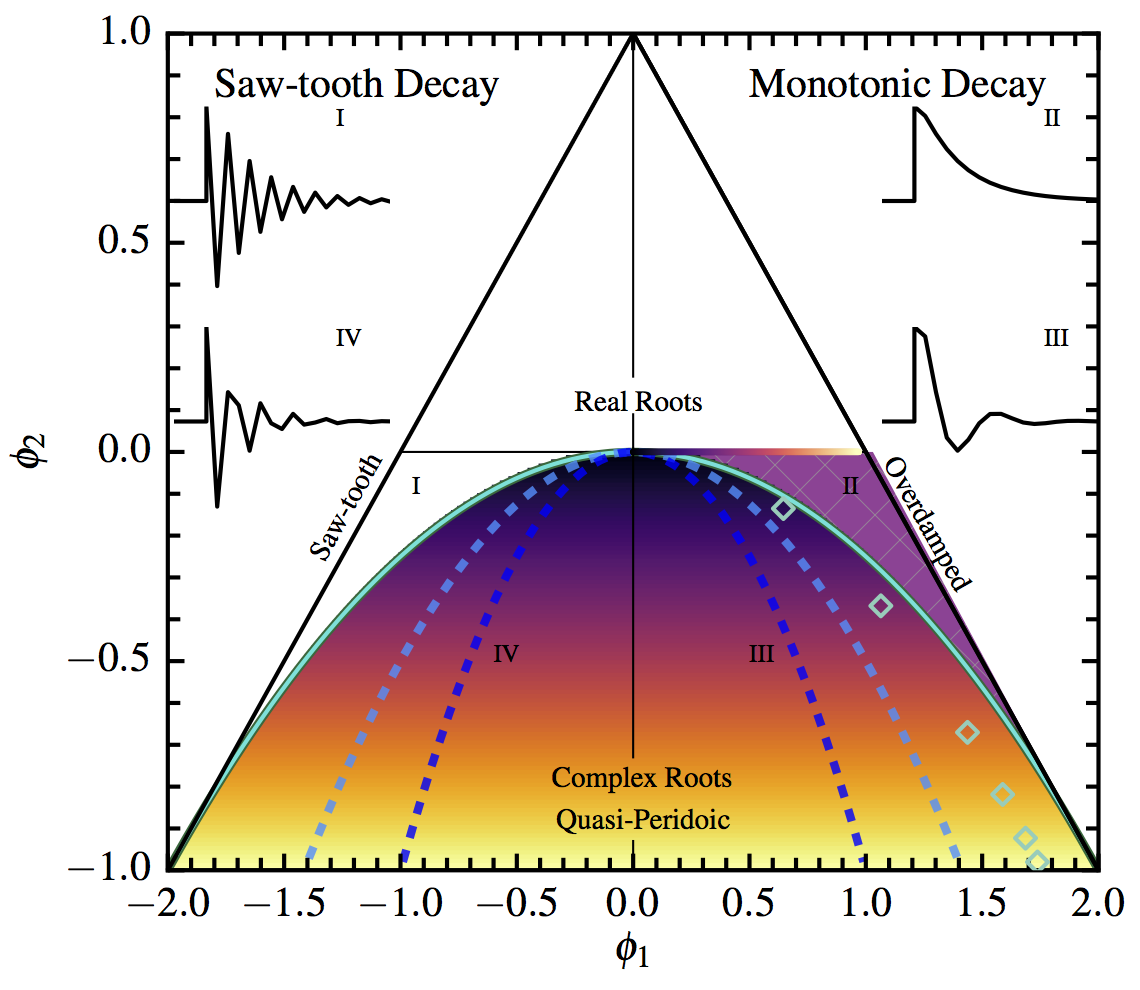}\\
    \caption{ARMA(2,1) Stability Triangle: a projection of CARMA space. Sawtooth decay occurs (inside the triangle) where $\phi_1$ and $\phi_2$ are both negative. Region I is the domain above the solid parabola and Region IV is the domain under the solid  parabola. Monotonic decay occurs (inside the triangle) where $\phi_1>0$ and $\phi_2<0$. 
    Region II (above the solid parabola) is over-damped and provides a rise timescale in addition to the decay timescale that is better known from DRW models ($\phi_1 > 0$ and $\phi_2 \rightarrow 0$). In Region III (under the solid parabola), CARMA roots are complex, where the imaginary component describes an average QPO frequency. Narrowing parabolas indicate increasing QPO frequencies. Diamonds sample the space see Fig\ref{Fig:Impulse} which we will discuss in Section 8. The color gradient (under the solid parabola and along the $\phi_2=0$ is the DRW subspace) tracks the decay timescale, increasing from dark to light.  We expect AGN to populate Regions II and III.}
	\label{fig:triangle}
\end{figure}

Thus far we have presented a limited range of ARMA processes because continuous-ARMA models are only defined in states of equilibrium, which correspond to limited ranges of coefficients, $\phi_1$ and $\phi_2$.  We illustrate the correspondence between discrete and continuous ARMA(2,1) models (the most complicated model that we will consider herein), using the stability triangle in Figure~\ref{fig:triangle}. The stability conditions, $\phi_1 +\phi_2 < 1$, $\phi_2 -\phi_1 <1$ and $-1<\phi_2 <1$  bound the triangle (\citealt{PhadakeWu}).  Regions {\em outside} of the triangle are non-stationary and run away with time.  Colored regions inside the triangle reveal the projection of the CARMA parameter space that overlaps the discrete space.

Guided by the stability triangle, we now move from discrete ARMA models (which correlate time lags) to CARMA models (which extract timescales of highly-correlated uneven data).   
A crucial difference between discrete and continuous systems involves changing notation from a differencing representation, $x_{i-p}$ (as in equation 2) to a differential form $d^{p}x$.  Thus, a second order CARMA process is described by
\begin{equation}
\label{eq:dho}
d^{2}x +\alpha_{1}d^{1}x+\alpha_{2}x = \beta_{0}\epsilon(t)+\beta_{1}d(\epsilon(t)). 
\end{equation}
The left-hand side of the equation, with  C-AR coefficients, $\alpha_1$ and $\alpha_2$,\footnote{Note that the coefficient of the second-order term is traditionally defined to be unity.} describes the auto-regressive part of the system.   The perturbation terms (i.e., the moving average part of the system), with coefficients $\beta_0$, $\beta_1$, are now on the right-hand side, indicative of their ``driving" nature, which we explore further in the next section.  This equation has the same differential form as a wave formula and is popularly referred to as a damped harmonic oscillator (DHO) model. 

Consistent with the standard manner of solving a differential equation, we set up the characteristic equation for the left-hand side as
\begin{equation}
r^2 + \alpha_1 r + \alpha_2 = 0,    
\end{equation}
to obtain the roots, $r_1$ and $r_2$.  These roots define the equation of motion, i.e., the solution to the CARMA process.

Second-order CARMA roots define two general domains of behavior: {\em complex} roots yield damped quasi-periodic behavior (Fig~\ref{fig:triangle}: below the solid outer parabola, regions III and IV) 
and {\em real} roots yield decaying behavior (Fig~\ref{fig:triangle}: colored region above the solid parabola, region II).  With these roots in hand, \citet{PhadakeWu} provide a way to visualize how CARMA roots map to ARMA coefficients with the following correspondence relationships:
$$\phi_1 = e^{r_1 dt} +e^{r_2 dt}$$
$$\phi_2 = -e^{(r_1 +r_2) dt}$$
where $r_1 $ and $ r_2$ are negative real roots or complex roots, which define the system dynamics. 

\subsection{Interpreting the Stability Triangle}

The stability triangle organizes categorical information about the response (dynamic behavior) of second order systems, which we unpack by highlighting differences and common attributes across four regions.  
In regions I and IV (left half of the triangle), the negative $\phi_1$ domain drives a sign flip between even and odd elements in a time series which creates a sawtooth response.  Regions II and III (right half of the triangle) exhibit monotonic decay. 
We expect astrophysical time series to predominantly occupy the right side of stability triangle where each decay timescale is monotonic (does not change sign).  

\begin{figure*}
	\plotone{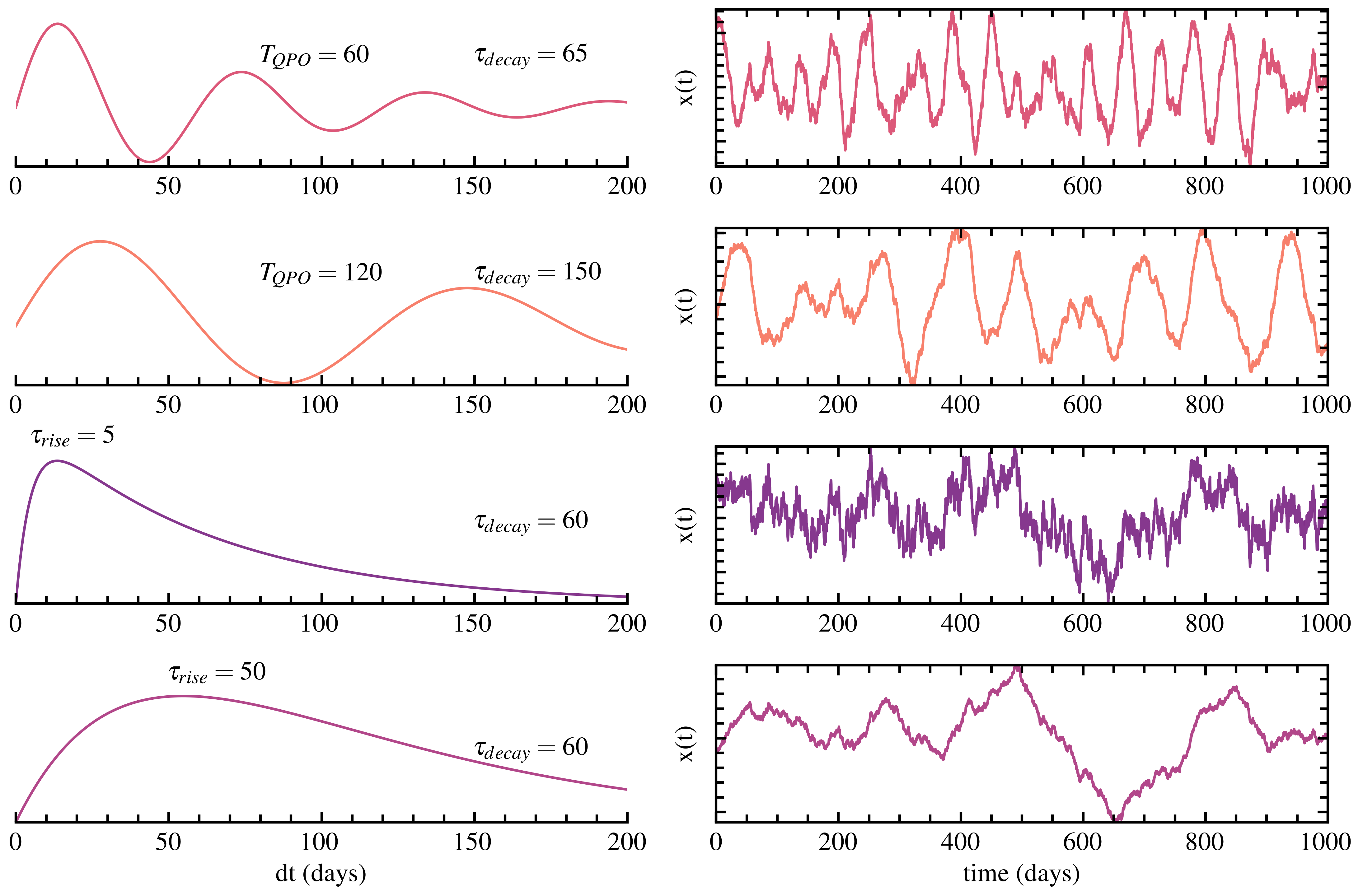}\\
    \caption{DHO processes from the quasi-periodic regime in the top two rows and over-damped regime in the bottom two rows. Each system response on the left acts on the same exact noise realization to produce the corresponding stochastic lightcurve on the right. The amplitude of these processes is determined by the perturbation coefficients in combination with these timescales. Colors map back to the approximate locations of the specified timescales in the stability triangle.}
	\label{fig:triangle_samples}
\end{figure*}

The complex CARMA roots, $r_1$ and $r_2$, provide timescales of variability that we can interpret in the stability triangle context.  The real parts of the roots are negative (inside the triangle) and have units of inverse time. 
In Figure~\ref{fig:triangle}, color gradients from dark to light correspond to decay timescales going from short to long respectively. 

The imaginary parts of the complex roots, $Im(r_i)$, define the frequency of a quasi-periodic oscillation (QPO), with period is $T_{QPO} = 2\pi$/Im$(r_i)$.  Each parabola, $\phi_{2} = -\frac{n}{4}\phi_{1}^2$ ($n = 1, 2, 4$ shown), traces a line of constant QPO frequency in the projected CARMA space. Moving up along a parabola, we traverse the color gradient in the figure, indicating a shorter {\em decay} timescale as we approach the darker end. 
The QPO frequency domain is continuous, but we illustrate three parabolas parameterized by $n$ which can be any positive, real number. To increase the QPO frequency of a process, we must increase the magnitude of $\phi_2$ relative to $\phi_1$ via $n$ to obtain a narrower parabola.  
Figure~\ref{fig:triangle_samples} shows examples of DHO processes in this quasi-periodic regime (the top 2 rows).  We show ``system response" curves with specified timescale combinations of $T_{QPO}$ and $\tau_{decay}$ in the left column.  Corresponding output lightcurves to each response are shown in the right column, using the exact same noise input for each.

The outer-most parabola borders a special case of critical/transition behavior (two {\em equal} negative-real roots).  Above the critical parabola, two {\em unique} negative-real roots (Region II) characterize over-damped behavior. Over-damped models are cousins of damped random walks.  They consist of a delay timescale over which perturbations grow plus the primary decay timescale $\tau_{decay}$ better known in DRWs. For over-damped models, we will refer to this additional timescale as the {\em rise timescale}, $\tau_{rise}$. When the rise timescale approaches zero we recover the DRW model ($\phi_2 = 0$, where the color gradient from dark to light also indicates short to long decay timescales).  Figure~\ref{fig:triangle_samples} also shows examples of DHO processes in the over-damped regimes (bottom two rows).  Again, we show ``system response" curves with specified timescale combinations of $\tau_{rise}$ and $\tau_{decay}$ in the left column and corresponding output lightcurves in the right column.   We dive deeper into system response and how it is used to extract characteristic timescales in Section~\ref{sec:response}.

We have now connected the correlation described by discrete AR and MA components (as in Figures~{\ref{fig:AR}} and \ref{fig:MAmany}) to the timescales provided by continuous models (as in Figure \ref{fig:triangle_samples}). 
Increasing the value of $\phi_1$ (when $\phi_2 \rightarrow 0$) corresponds to increasing the DRW decay timescale, i.e., \textit{less efficient dissipation} of energetic perturbations. Increasing the magnitude of $\phi_2$ relative to $\phi_1$ roughly corresponds to increasing the secondary characteristic timescale, either as an average QPO periodicity (complex roots) or as a rise timescale (real roots). When the rise timescale approaches zero we truly have a DRW.  

In Sec~\ref{sec:Nutshell}, we expose the mathematics behind the graphical representations of CARMA processes that we have previewed from the stability triangle in Figure~\ref{fig:triangle_samples}. We expect AGN lightcurves that are well modelled by DRWs to cluster with very short rise timescales (close to zero days).  Lightcurves that are better modelled by DHOs should lie in the lower right corner of the stability triangle (in both Regions II and III). Our goal with the stability triangle is to gain an intuition for navigating the continuous model parameter space which is a much broader landscape. It is useful to understand the stability triangle as a phase diagram, where the simplest behavior approaches a DRW, more complex behavior appears in the form of a rise timescale (the over-damped regime), transition behavior appears as a highly damped QPO, and finally a relatively high frequency QPO may be the signature of a secondary physical process that becomes dominant.

\section{Perturbation in Continuous Systems}
This section delves into the interaction of deterministic (C-AR) and stochastic (C-MA) components in a CARMA model as we move away from discrete processes. A stock price influenced by demand is a discrete process that changes in time as a result of transactions of selling and buying.  In contrast, the swinging of a pendulum or an unemployment index do not cease to exist between measurement intervals, but rather they evolve continuously in time. Continuous systems exhibit responses to disturbances rather than instantaneous redirection upon impact; in physical systems this is the property known as {\em inertia}. The latter set of systems are also subject to \enquote{continuous} perturbation mechanisms that drive changes to the system stochastically rather than deterministically. 

\begin{figure} 
	\plotone{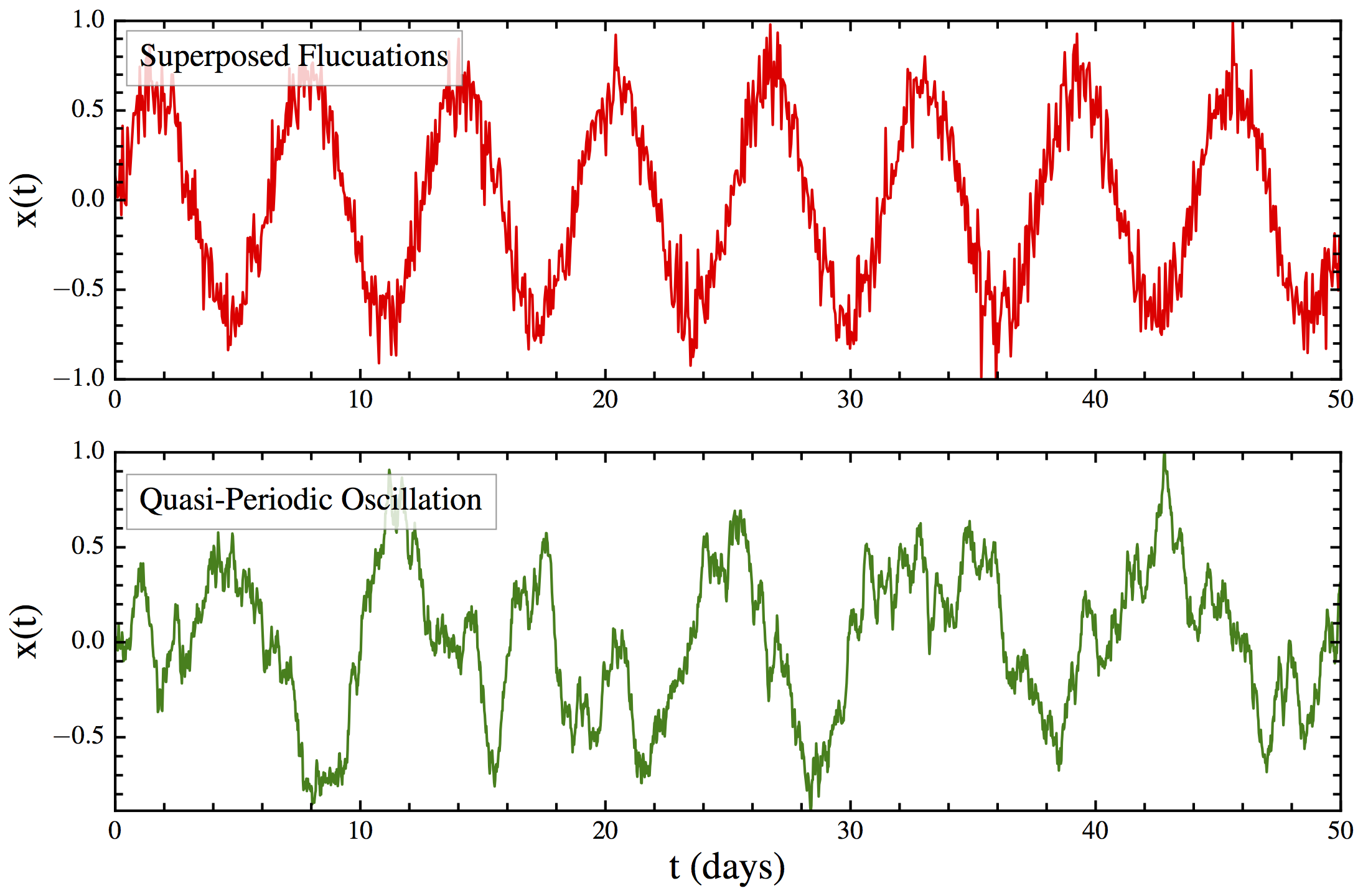}
    \caption{Superposed fluctuations: a pendulum with measurement error (top), and  a pendulum driven by perturbations showing quasi-periodicity (bottom). The origin of quasi-periodic (as opposed to truly periodic) behavior arises from a stochastic process acting on an inertial system.}
    \label{Fig:qpo}
\end{figure}

Suppose a pendulum is swinging back and forth with a constant amplitude.  We observe small measurement errors during its harmonic motion about its equilibrium position.  A graph of the measured position as a function of time remains visibly periodic as in Figure~\ref{Fig:qpo} (top).  If the observation errors are large (compared to the amplitude of the periodic component), the underlying periodic behavior is obscured by \enquote{superposed fluctuations}.  Now, consider a pendulum that initially sits at rest in its equilibrium position when a group of children begin pelting the pendulum with pebbles. The motion is subject to a true disturbance (as opposed to measurement error) and the effect on the graph reflects an entirely different character.   The graph of the pendulum's motion remains intrinsically harmonic but the amplitude and phase vary continually as illustrated in the bottom panel of Figure~\ref{Fig:qpo}. \citet{Yule1927} proposed such a thought experiment to highlight the effect of a randomly driven system having inertia: the effect is not simply periodicity with superposed fluctuations but quasi-periodicity.

Yule's pendulum demonstrates how quasi-periodic (as opposed to truly periodic) behavior arises from a stochastic process acting on an inertial system. In Figure~\ref{Fig:qpo}, we illustrate the difference between superposed fluctuations (measurement error) compared to intrinsic perturbation and we recognize that Yule was describing quasi-periodicity.

In second order CARMA processes (Equation 7), differential terms (on the LHS) model viscous forces that dissipate energy (maintaining equilibrium) and inertia which similarly resists changes in motion and direction.  The LHS can deterministically describe periodic motion in the trivial case (RHS = 0).  
A non-trivial case specifies a driving mechanism for example, pebbles kicking the pendulum (RHS is stochastic). Quasi-periodic behavior arises only with the right combination of ``viscous" and perturbation parameters.

The DHO model characterizes noise with two perturbation parameters, one that measures amplitude and another that measures covariance.
A continuous white noise process is generated by a Gaussian distribution with constant variance for all lengths of the interval $dt$.
If the variance of the noise process exhibits a rate of change, as the interval length is varied, the perturbation function has some correlation or covariance. 
The rate of change of the variance is captured in the MA coefficient. 
A continuous MA process ultimately governs the variance of the noise-generating process.
In the most complex process that we consider,  CARMA(2,1), a perturbation function with two coefficients $\beta_0$ and $\beta_1$ describes noise drawn from a density of all possible random values. The larger the ratio of MA coefficients $\frac{\beta_1}{\beta_0} > 1$, the more erratic (bluer) the behavior of the time series. 
We summarize the minimum necessary mathematics of continuous stochastic processes in the next section.

\section{CARMA in a Nutshell}
\label{sec:Nutshell}

\begin{figure*} 
	\plotone{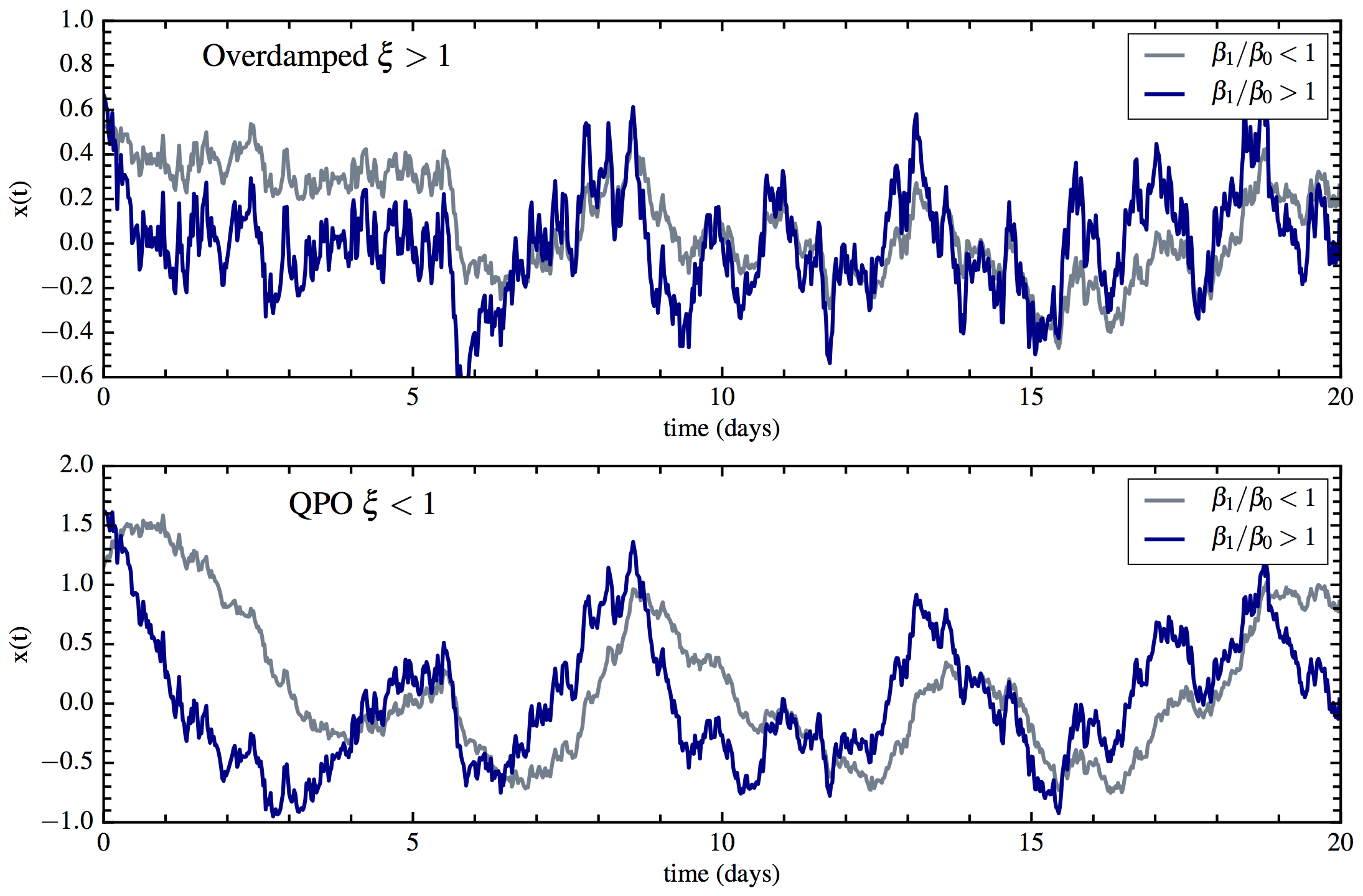}
    \caption{Illustration of damping and perturbation effects. Over-damped models are shown in the top panel and QPO (under-damped) models are shown in bottom panel. Grey curves feature a white-noise-dominated perturbation process, $\beta_1/\beta_0 < 1$, and blue curves features a blue-noise-dominated, $\beta_1/\beta_0 > 1$, process.  The perturbation characteristics of the mock lightcurves are subtle in the time domain but they become more apparent in frequency space (see section 10.)}
    \label{Fig:DampingFactor}
\end{figure*}

The simplest of CARMA models is the damped random walk (DRW or CARMA(1,0)). The  DRW can be expressed in differential form as
\begin{equation}
d^1x +\alpha_1x(t) = \beta_0 dW(t),
\label{eq:eq9}
\end{equation}
where $\alpha_1$ is the C-AR coefficient and $\beta_0$ is the coefficient of random perturbations. 
The perturbation function on the RHS is also redefined using Wiener increments $dW$ rather than the random variable $\epsilon(t)$ as in previous equations, however both notations are equivalent by Ito's theorem (\citealt{Ito}).
The solution to Equation~\ref{eq:eq9} is a ``forecasting equation", which can be expressed as a so-called ``state space solution": 
\begin{equation}
x(t+\delta t) = e^{\alpha_1 \delta t} x(t)+\beta_0 \int_0^{\delta t} e^{\alpha_1(\delta t-s)} dW_s.
\end{equation}
Here $x(t)$ represents the current state of the observed system, $\alpha_1 = |r_1|$ 
in the case of a first order process, $\beta_0$ equals the standard deviation of the process,
and finally, the integral over Wiener increments $dW$ fully describes the density function of white noise perturbations (\citealt{Jones1993}, \citealt{brockwell2013}, \citealt{Kelly14}). 
This model has of two degrees of freedom:  a decay or relaxation timescale, $\tau$, and a limiting amplitude of variability $\beta_0$.   
We obtain the decay timescale from the auto-regressive coefficient $\tau =-\frac{1}{r_1} = \frac{1}{\alpha_1}$.  The amplitude $\beta_0 $ is coupled to the decay timescale $\propto \sqrt{c \tau/2}$, where the constant $c$ is analogous to the diffusion constant in an Ornstien-Uhlenbeck (1D-diffusion) process (\citealt{brownian}).  Given that the perturbation function is a white noise process, $\beta_0$ takes the role of an ``amplitude-like" parameter when multiplied by the density function.  
The density function, is also referred to as a univariate Gaussian, drift-less Wiener process or Ito integral (\citealt{brownian}, \citealt{Vish2017}).

CARMA(2,1), a second order differential equation driven by a continuous MA(1) process,  was introduced in Equation~\ref{eq:dho}; however, it is useful to present it in a form re-written with a change of variables:
\begin{equation}
\label{dhodiffeq}
d^{2}x +2\xi \omega_0 d^{1}x+\omega_0^{2}x = \beta_{0}dW(t)+\beta_{1}d(dW(t)).
\end{equation}
The C-AR coefficients are now $\alpha_1 = 2\xi \omega_0$ and $\alpha_2 = \omega_0^2$.
$\xi$ is the damping ratio of the oscillator and $\omega_0$ is the natural QPO frequency. When the damping ratio, $\xi$, is greater than 1 (over-damped), the system experiences delays in the growth of perturbations that drive the state $x$ away from equilibrium and a delay as the system decays back to equilibrium. 
In this case, the roots $r_1$ and $r_2$ are negative-real and describe both a rise timescale $\tau_1 = \frac{-1}{r_1}$ and a decay timescale $\tau_2 = \frac{-1}{r_2}$.

DHO behavior is obtained by setting up a characteristic equation and solving for two roots,
\begin{equation}
 x(t+\delta t) = e^{r_1 \delta t} x(t) + e^{r_2\delta t} x(t) +\int_0^{\delta t} e^{{\bf A}(\delta t-s)} {\bf B} dW_s
\end{equation}
The density function of the continuous MA(1) process now involves the matrices 
{\bf A } = 
$$\begin{bmatrix}
    -\alpha_1     & 1 \\
     -\alpha_2 & 0\\
\end{bmatrix}$$ and {\bf B} = [$\beta_1,\beta_0 $]$^T$ (\citealt{Kelly14} and references therein).  When parameters are specified (data is fitted) it defines a multivariate Gaussian $\mathcal{N}(\bf{0},\bf{C})$ having a zero mean vector and a disturbance covariance matrix ($\bf{C}(w_i, w_j) = C_{i,j}$). 
We show this notation for full transparency, however, it is easier to think of both $\beta_0$ and $\beta_1$ terms as white noise and blue noise terms, respectively, as discussed previously in Section 3. For more information we refer the reader to \citet{Jones1993} and \citet{Vish2017}.  

For underdamped systems ($\xi <1 $), the roots are complex conjugates with $r_1 = (\frac{-1}{\tau}+\omega_0 i)$ and $r_2 = (\frac{-1}{\tau}-\omega_0 i)$. The damped oscillator frequency is $\omega_d = \omega_0\sqrt{1-\xi^2}$. Applying Euler's formula, the DHO state equation may be rewritten by expanding the complex exponentials into sines and cosines, rendering a full real-space solution of the form:
$$x(t+\delta t) =  x(t)e^{\frac{ -\delta t}{\tau}}sin(\omega_d t +\phi) + perturbations.$$
The effect of perturbations on the phase $\phi$ of the harmonic term is to cause ``quasi"- periodic behavior.  Examples of a DHO with parameters on each side of critical damping are shown in Fig~\ref{Fig:DampingFactor} (over-damped (top) and under-damped bottom). A QPO is more apparent in a light curve when the damping ratio $\xi < 0.5$.  When the roots are both real, quantities like the damping ratio, $\xi$, and $\omega_0$ lose their significance.

\section{Impulse Response}
\label{sec:response}
\begin{figure*}
    \plottwo{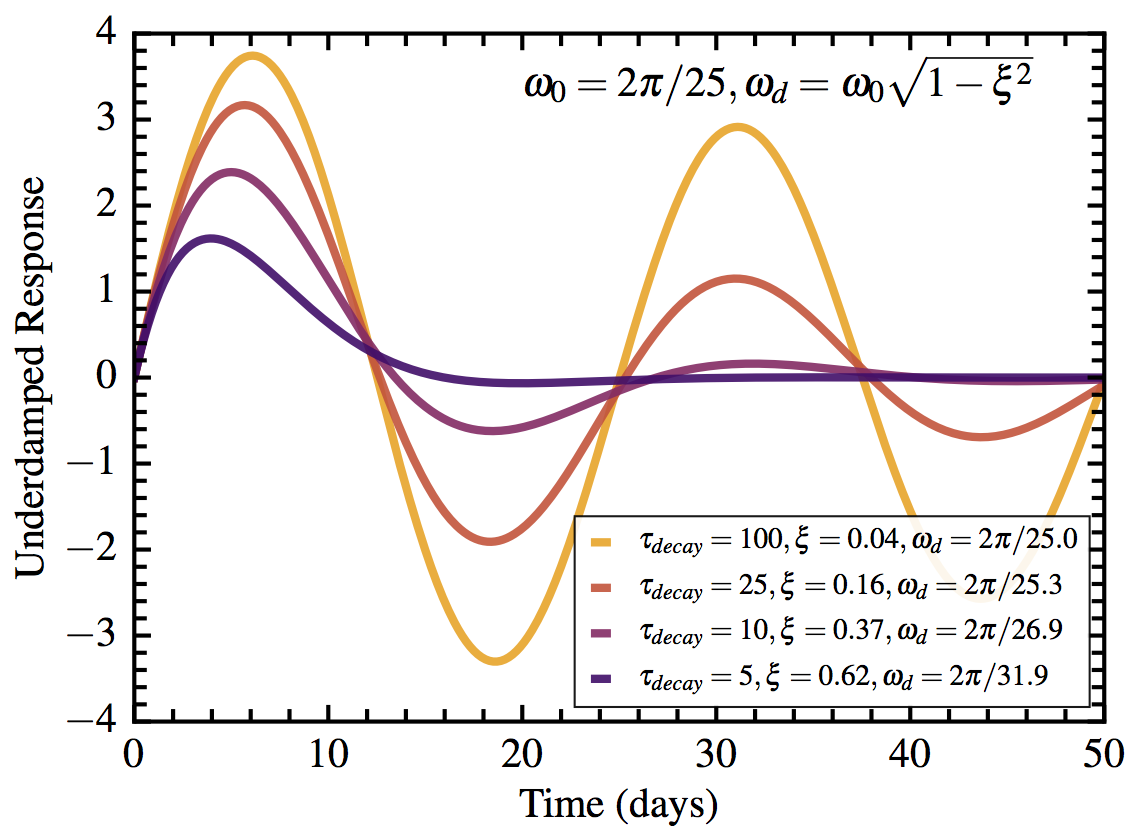}{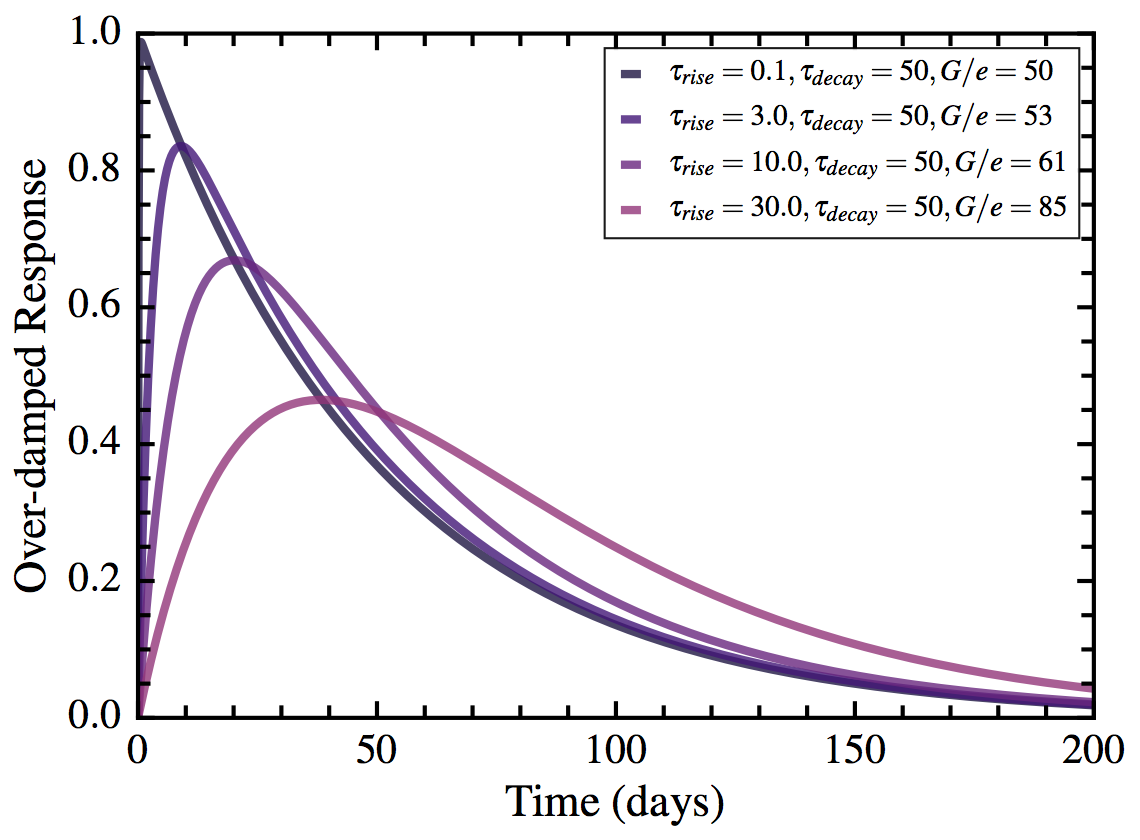}
   	\caption{ QPO response (left) and over-damped response (right). Each curve on the left panel is the response of a process sampled from the diamonds in the stability triangle (Figure~\ref{fig:triangle}). In the QPO-response, the frequency is set to a constant and the decay timescale $\tau_{decay}$ is increased from dark to light tones.  In the case of the over-damped response, the decay timescale is held constant and the rise timescale is increased from dark to lighter tones. For both responses, darker tones are closer to DRWs }
	\label{Fig:Impulse}
\end{figure*}

In a similar manner as we did for ARMA processes, we gain insight into how CARMA processes (and the DHO model in particular) work by examining how the system responds to a delta-function impulse.  Specifically how CARMA processes represent {\em causal} time series. 
Mathematically, the impulse response function is a Green's function of the differential equation.
This impulse response function, which is generally asymmetric in time (as shown previously in Figure~\ref{fig:triangle_samples}), describes how a disturbance to the system at one time propagates into the future state of the system. Statistical characterization of the time series using second-order statistics (e.g., PSD's and structure functions, which will be discussed later) neglects this time-ordering---thus losing information that can be critical to a physical interpretation.   The CARMA description captures not only the statistical properties of the driving disturbances, but also the characteristic response in time to those disturbances. As discussed above for the DHO model, this approach yields timescales for the rise and subsequent decay of noise-driven perturbations.

Yule's pendulum demonstrates how crucial Green's functions are to developing our understanding and application of CARMA processes.
Previously, we examined this physical example to discuss perturbation and quasi-periodicity, now we refocus on the motion of the pendulum itself.
The  pendulum's response to a single pulse is the Green's function $G$.  The actual motion of the pendulum is the linear superposition of multiple pulse responses.  In the infinitesimal limit, the pendulum would be perturbed by a continuous spectrum of force pulses, represented as delta functions $\delta (t)$, with the effect on the pendulum's state characterized by the same Green's function.

The Green's function of a CARMA(p,q) process can be derived by describing the RHS of the CARMA differential equation as a Dirac delta function $\delta (t)$,
\begin{equation}
\frac{d^pG}{dt^p}+ \alpha_1 \frac{d^{p-1}G}{dt^{p-1}} +...+\alpha_p G = \delta(t).
\end{equation}
The order, $p$, of the differential equation describing the CARMA process dictates the complexity of the impulse-response of the system, which is given by
\begin{equation}
\label{eq:greens}
G(t) = \Sigma^{p}_{k} c_{k} e^{r_{k} t}.
\end{equation}
Here the constants, $c_k$, for $1\leqslant k \leqslant p $ are defined by a matrix involving the roots of the process discussed in the previous section,

\begin{equation}
\begin{bmatrix}
r_{1}        & r_{2} \\
1	& 1 
\end{bmatrix}
\begin{bmatrix}
c_1 \\
c_2 \\
\end{bmatrix}
=
\begin{bmatrix}
1 \\
0 \\
\end{bmatrix}.
\end{equation}
This matrix is essentially a summary of boundary conditions for $p = 2$ derivatives of the CARMA process. The interested reader is referred to Appendix B of \citet{Vish2017} for the general form of the matrix.  The boundary conditions for a second order process allow us to solve for the DHO Green's function.
Thus, we expand the sum in Equation~\ref{eq:greens}  and solve the for coefficients $c_1$ and $c_2$ to obtain 
\begin{equation}
G(t) = \frac{e^{r_1 t}-e^{r_2 t}}{r_{1}-r_{2}},    
\end{equation}
where $r_1$ and $r_2$ are again the roots of the characteristic equation.

The shape of the DHO's Green's function has a different interpretation for the two regimes of behavior that were discussed in the ARMA triangle  (Figure~\ref{fig:triangle_samples}).
When the roots of a DHO are complex conjugates (left panel), $G(t)$ reduces to 
$$G(t) = \frac{1}{\omega_d}e^{\frac{ -1}{\tau} t} sin(\omega_d t)$$
where the damped natural frequency is $\omega_d = \omega_0\sqrt{1-\xi^2}.$
In this case, the Green's function contains a harmonic term as illustrated in Figure~\ref{Fig:Impulse} (left).  A lightcurve with such a Green's Function exhibits multiple peaks and troughs of variable width. 
In the over-damped case (right panel), the system's viscous forces are highly efficient and they contribute a delayed response to perturbations. Since the time delay describes the growth of perturbations (how the lightcurve moves away from the mean) we refer to it as the rise timescale.  Figure~\ref{Fig:Impulse} (right) shows that as rise timescales approach zero, the DHO model becomes degenerate with the DRW response $G_{DRW}(t) = e^{r_1 t}$. 
This means that a lightcurve with a 5 day rise timescale is closer to a DRW than a lightcurve with a rise timescale of 50 days. The PSD of such lightcurves (with relatively short rise timescales), 
would be characterized by high frequency deviations from the DRW slope ($\nu^{-2}$) but we would be unable to interpret exactly how the drop or excess in power manifests in the time domain (further discussion in Section~\ref{sec:PSD}).

\section{Auto-Correlation}
\label{sec:ACF}
\subsection{The ACVF}

Related to Green's Functions (and PSDs and structure functions)
are auto-covariance (ACVF) functions, which are fundamental to an astromoner's time-series tool kit.  
The sample auto-covariance of a time series is
\begin{equation}
\label{eq:ACVF}
ACVF(\Delta t) = \frac{1}{n}\sum^{n-j}_{i = 1, j = 0}{(x_{i-j}-\overline{x})(x_{i}-\overline{x})}
\end{equation}   
where $x_{i-j}$ is the state at a time lag $\Delta t$, and the value of the ACVF at lag $\Delta t = 0$ (where $i=j$)  is equivalent to the variance of the data. For all $\Delta t > 0$, we get the averaged covariance or correlation of binned pairs of points separated by time lags of increasing length. 

The auto-covariance (ACVF) function of a time series with itself gives a measure of the correlation strength of an evolving process with its own fluctuation history as a function of time lag. 
The correlation strength shown by the ACVF is a very informative guide for model selection. In the case of stochastic or noisy data, the ACVF informs our model building or prediction algorithms by revealing if there is any persistent time dependence or correlation structure.  

\subsection{The ACVF of the DRW and DHO}
The auto-regressive parameters of the DRW and DHO may be used to estimate an analytic ACVF. In Figure~\ref{Fig:ACF}, we show the graphical representations of the analytic ACVF of a DRW and multiple DHOs with varying roots.
The ACVF of a DRW is 
\begin{equation}
ACVF_{DRW}(\Delta t) = \sigma_{process}^2e^{\frac{-\Delta t}{\tau}}=\beta_0^2 e^{r_1\Delta t}
\end{equation}  
Normalizing the ACVF by the variance produces a related function, the auto-correlation function (ACF), which offers another characterization of the data: 
\begin{equation}
\label{eq:acfdrw}
ACF_{DRW}(\Delta t) = e^{r_1\Delta t}.
\end{equation} 

The ACVF of the DHO or CARMA(2,1) model is instead
\begin{equation}
\label{eq:acfdho}
ACVF_{DHO}(\Delta t) = C_0 \bigg(C_1 e^{r_1 \Delta t} + C_2 e^{r_2 \Delta t}\bigg)
\end{equation}  
with weights (given two unique real roots or a complex conjugate pair \footnote{The symbol $^*$ is the complex conjugate and $\Re(r_i)$ is the real part of the complex value.}), 
$$C_0 = -\frac{ \beta_1 }{2 (r_2-r_1)(r_2^* + r_1)} $$
$$C_1 = \frac{r_1}{\Re(r_1)}(\beta_0-\beta_1 r_1),$$
$$C_2 =  \frac{r_2}{\Re(r_2)}(\beta_0-\beta_1 r_2).$$
The ACVF of a CARMA process described in Eq~\ref{eq:acfdho} has a very similar mathematical form to the Green's function shown in Eq.~\ref{eq:greens}. It is either a weighted sum of exponentials (for real roots) or a weighted sum of exponentially damped harmonics (for complex roots); see \citet{Kelly14} for the generic ACVF formula.  

The ACVF does not capture the same time-ordered information as the Green's function since it is a function of time lag not time;
however, its advantage is that it takes into account the perturbation terms, $\beta_0$ and $\beta_1$.  We can think of Green's functions as tools providing a system of solutions fitted to boundary conditions. The ACVF, as specificed by the roots (or eigen values) of both the LHS \textit{and} the RHS of a CARMA differential equation is a unique solution.  
The ACVF can therefore be used 
to approximate a CARMA process since it accounts for C-AR and C-MA parameters (see \citet{Celerite} for an alternate form of the DHO ACVF).  Note that there are two representations of C-MA coefficients in the astronomical literature: one notation is expressed in units of the standard deviation $\sigma$ ($\sigma$ is factored out of the RHS) therefore $\beta_q = 1$ is fixed (\citealt{Kelly14}), and in the other notation the standard deviation is not factored out and is therefore equal to the highest C-MA coefficient $\beta_q = \sigma$ (\citealt{Vish2017}).  We use the latter representation in all our equations involving $\beta_0$ and $\beta_1$.

\subsection{Non-stationarity, Noise and Sampling Effects }
The definition of the ACVF implicitly requires that the time series have a constant mean and variance for the shape of the ACF or ACVF to be meaningful. 
In statistical literature, this combination of properties is referred to as a \enquote{stationary} process: any segment of the light curve looks (statistically) identical to any other segment of the same length.    A strong indicator of non-stationarity is if the ACF of the process does not decay.
We might also observe a steep linear trend or runaway as examples of non-stationary behavior.  In astrophysical applications, our systems can be treated as approximately stationary through the appropriate choice of a window function in time. It is important for astronomers to confront the fact that non-stationarity affects all second order statistics like PSDs, structure functions, and CARMA models. The variability statistics of non-stationary or short lightcurves (relative to intrinsically long timescales) may still be effective for classification since they will operate as indicators of non-stationarity (for more relevant empirical work regarding length systematics see \citealt{KozlowskiLength}). 

\begin{figure} 
	\plotone{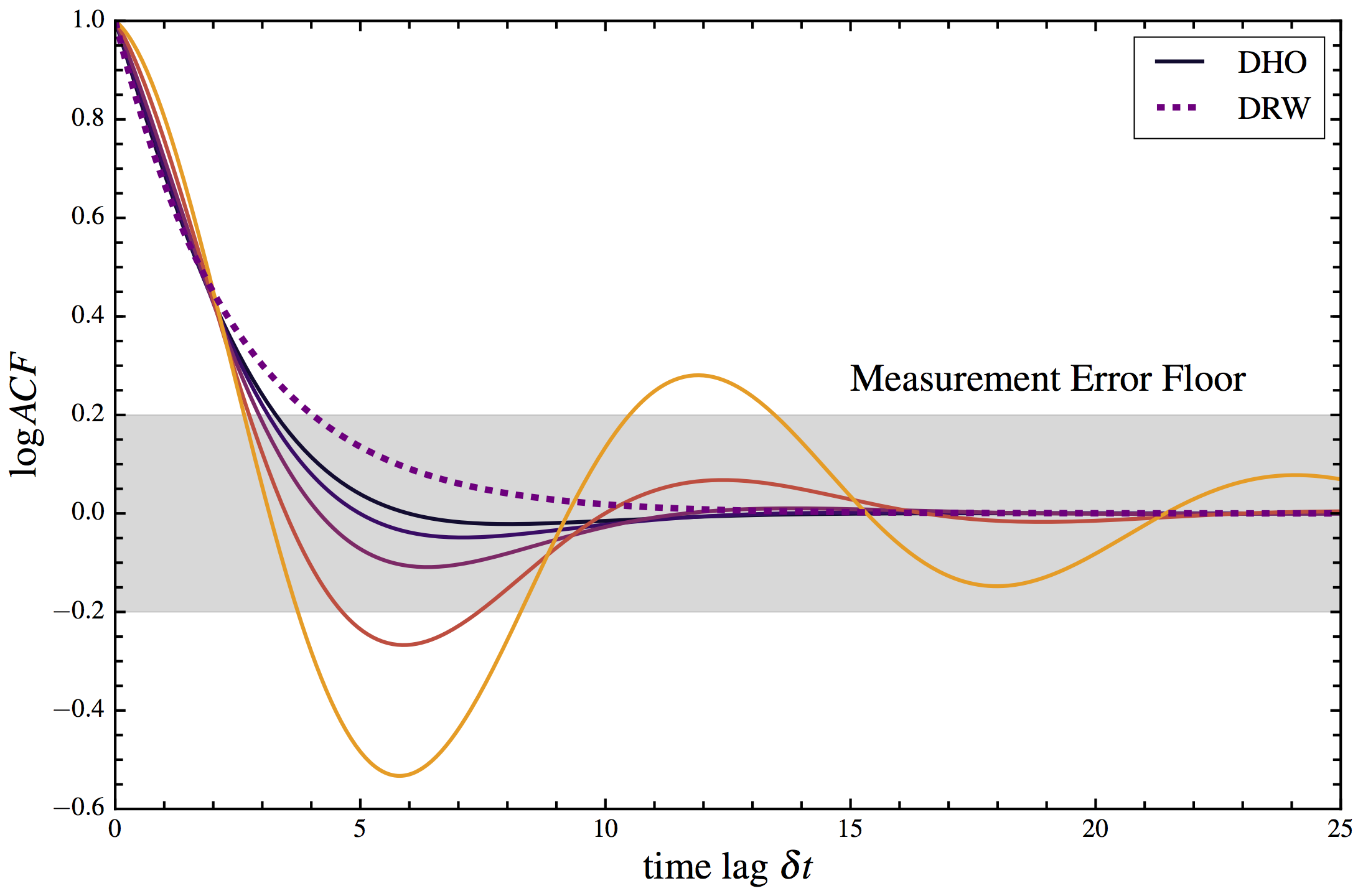}
    \caption{The analytic ACF of a DRW (dashed purple) and DHO with relaxed damping $\xi$ from over-damped (dark, solid) to under-damped (light, solid). Higher order correlation structure may be obscured in the error floor (gray band), thus we potentially lose the ability to resolve a rise timescale and the data would show satisfactory agreement with a DRW model. If the model appropriately represents the data, then the analytic curves should follow the ACF of the data, thus the ACF and ACVF are excellent diagnostics for model selection.}
    \label{Fig:ACF}
\end{figure}

Measurement noise can also affect the estimate of the lightcurve variance and the resolution of the true amplitude of variability. Since the amplitude can be smaller at shorter time lags, a large noise floor can obscure true variability as diagrammed in Figure~\ref{Fig:ACF}. 
If the measurement error is large, the choice of model would necessarily be simpler because the data does not resolve higher order structure.  For example, a DHO simulated with large errors may appear like a DRW as shown by the ACFs in Figure~\ref{Fig:ACF}: the
same correlation is well described by multiple models, yet they all diverge below the noise floor. If the ACVF of a lightcurve shows correlation structure above the noise floor, then a second order model may provide more meaningful parameters. 

\begin{figure}
    \plotone{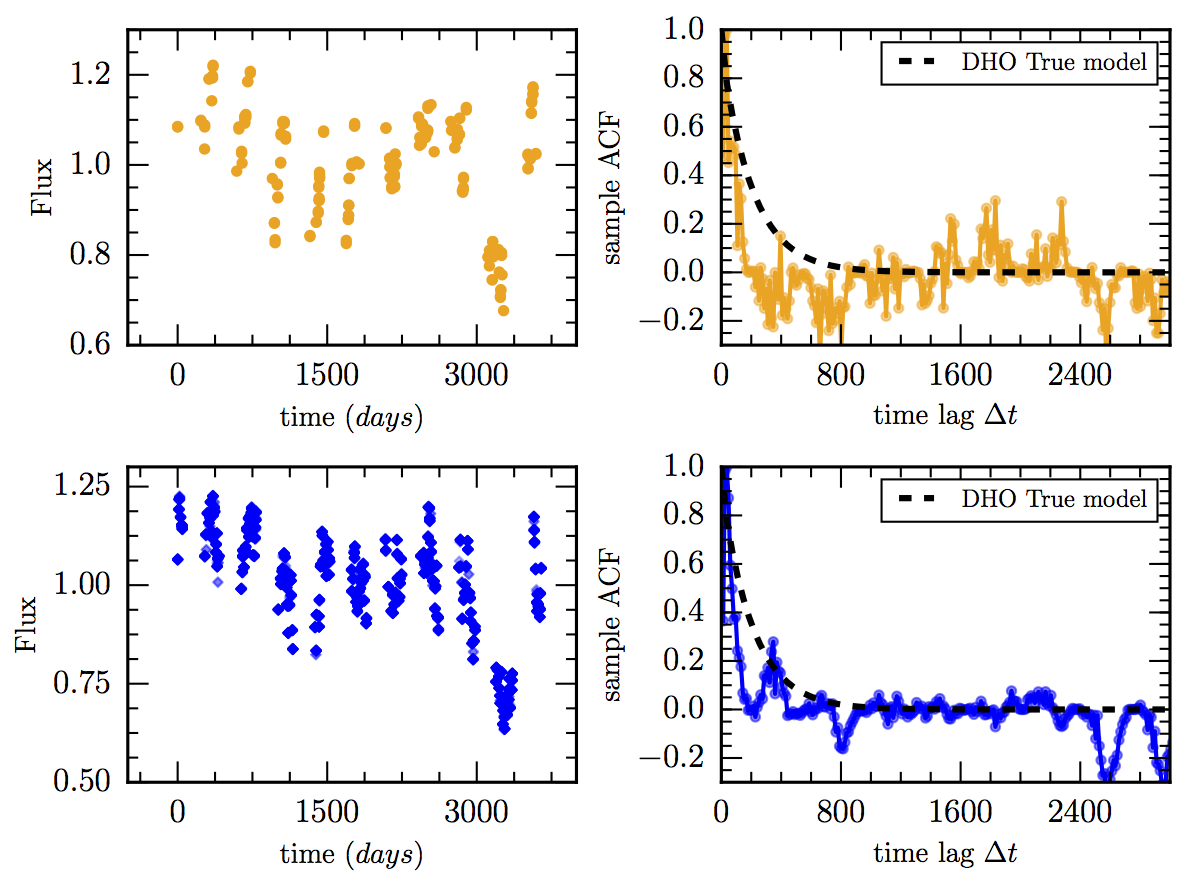}
    \caption{Sample ACF of a DHO mock lightcurve observed-like LSST-WFD (in orange) and LSST-DDF cadences (in blue). The dashed lines in black show how well the ``true" DHO model follows correlation measured by the sample ACF at short lags. Both ACFs are computed with the same binning, however the DDF cadence ACF has better agreement with the true ACF. The initial drop in the WFD and DDF ACFs are due to 6 month observability gaps.}
    \label{Fig:LSSTACF}
\end{figure}

When a process is under sampled, its time series appears uncorrelated even at the shortest resolved time lags and the ACF decays extremely rapidly to zero due to sparse sampling. 
A sparse cadence simply means that binned time lags have non-equal density of observation pairs, therefore parts of the ACF will be poorly resolved or entirely unresolved.  Figure~\ref{Fig:LSSTACF} shows the sample ACF of a mock AGN lightcurve with sampling like LSST's Wide Fast Deep (WFD) in orange and Deep Drilling Fields (DDF) in blue (for the r-band). The DDF cadence will capture more data at high frequency (short timescale), however, it is unclear if timescales inside observability gaps can be recovered reliably. LSST's WFD cadence is sufficient to recover truth parameters of mock the lightcurve with the max error being about a 50\% under-estimation of $\alpha_2$ and $\beta_0$, while $\alpha_1$ and $\beta_1$ had less than 1\% error. This example worked well for these parameters however, LSST cadences may not behave well with all combinations of DHO parameters. Thus, this requires further investigation. 

The ACVF is a powerful diagnostic for model selection when considering the DRW or the DHO and for determining whether a gappy time series captures sufficient correlated behavior to assume a model more complex than white noise in the first place. The sample ACVF and analytic ACVF should be overplotted (as demonstrated in Figure~\ref{Fig:LSSTACF}) and compared to determine if higher order correlation structure merits a more complex model. 

\section{Power Spectra}
\label{sec:PSD}

\begin{figure} 
	\plotone{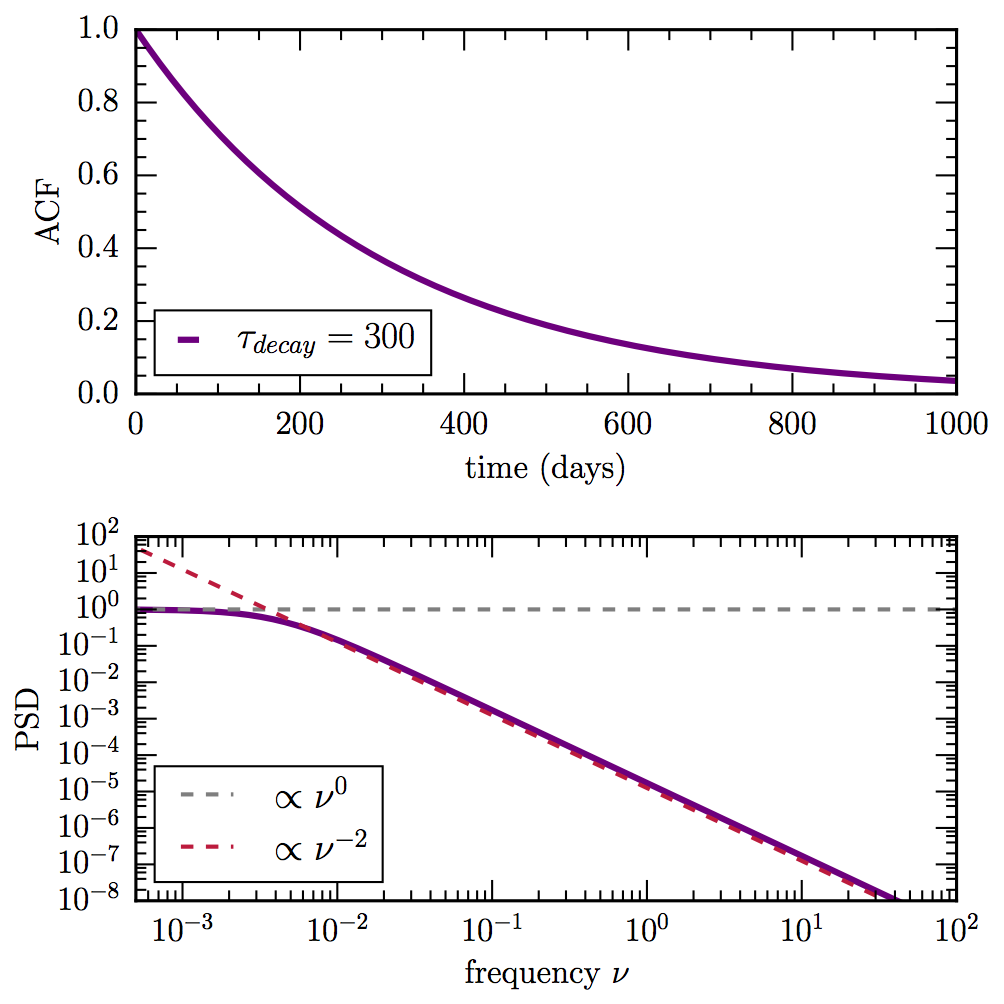}
    \caption{The ACF and PSD Fourier transform pair. A broken power law is one of the simplest models used to describe AGN variability, and it has the same form as the PSD of a DRW process which has a fixed amplitude-frequency scaling of $\nu^{-2}$.}
        \label{Fig:PSD1}
\end{figure}
While the ACVF provides a complexity diagnostic in time space, the power spectral density (PSD) provides a complementary diagnostic in frequency space.  A Fourier transform of the ACVF, $\gamma(\tau)$, is equivalent to the PSD
\begin{equation}
\label{eq:acfPSD}
S_{xx}(\omega) = |\int^{\infty}_{-\infty}\gamma(\tau)e^{-i\omega \tau}d\tau|^2.     
\end{equation}
Power-spectral analysis decomposes the power in a time series into a linear combination of complex exponentials (harmonics) that are characterized by an amplitude and frequency.  Equation~\ref{eq:acfPSD} represents the scaling of squared variability amplitude as a function of frequency. 

The simplest model used to describe AGN variability corresponds to a power law in frequency space (e.g., \citealt{Schmidt2010}).  
Typical AGN PSD estimates (which are historically poorly resolved) exhibit pink or red spectra, $$PSD \propto \nu^{-2},$$ and a break frequency at which the spectrum turns over to a flat slope at low frequencies (see Fig.~\ref{Fig:PSD1}).
The negative slope indicates that amplitude decreases with increasing frequency.  The flat turnover corresponds to (uncorrelated) white noise---signifying that resolved fluctuations are independent at these timescales (or maximal length scales) in AGN accretion disks.
Astronomers can extract timescales of interest when the empirical spectrum deviates from the theoretical fixed-slope scaling relationship.  These deviations may take the form of an inflection point leading into a slope flattening (decorrelation), steepening or a  Lorentzian (QPO signal). 

Thus, in this section we discuss the assumptions of adopting a single-sloped power law (CARMA(1,0)) versus a multi-sloped power law (e.g., CARMA(2,1), as illustrated in Figure~\ref{Fig:PSD}) and how it relates to multiscale analysis of AGN variability in the frequency domain. Deviations from a single slope reveal characteristic timescales that may point to multiple mechanisms for generating or dissipating energy at different scales. We explain such timescale interpretation by examining the analytic functions of DRW and DHO PSDs.

\subsection{PSD of DRWs and DHOs}

\begin{figure*} 
	\plottwo{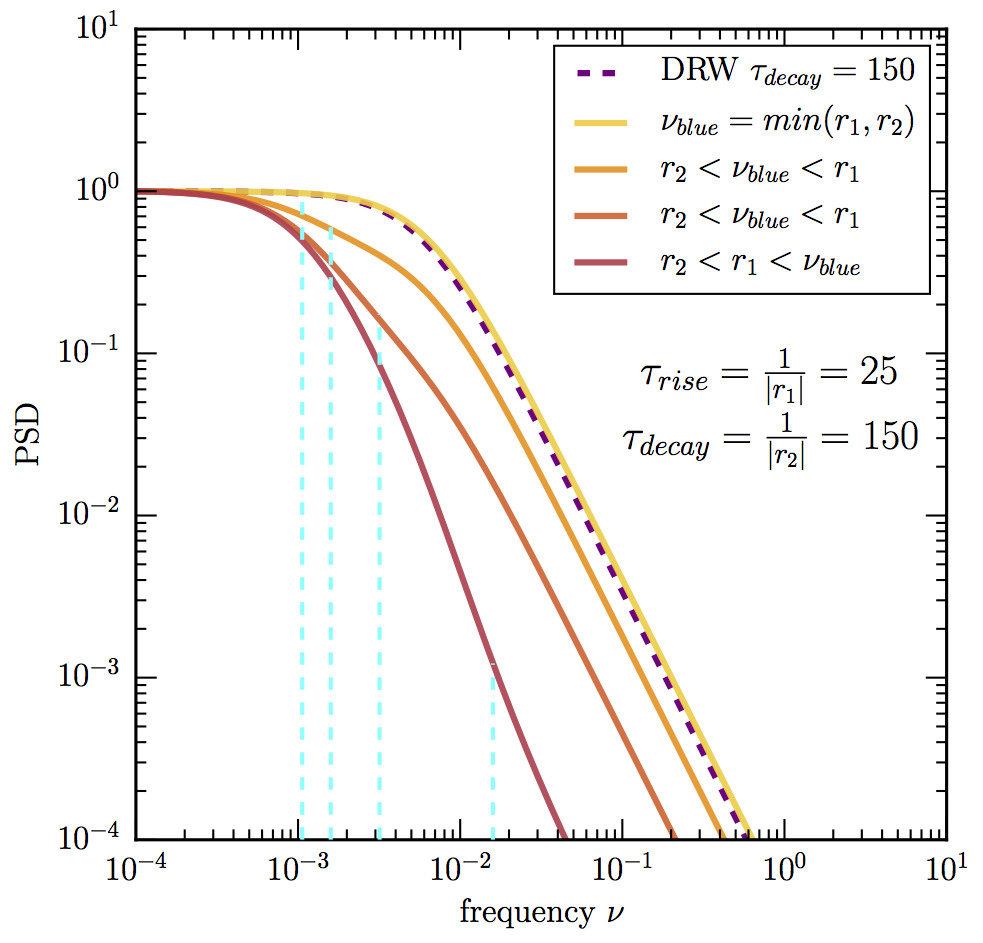}{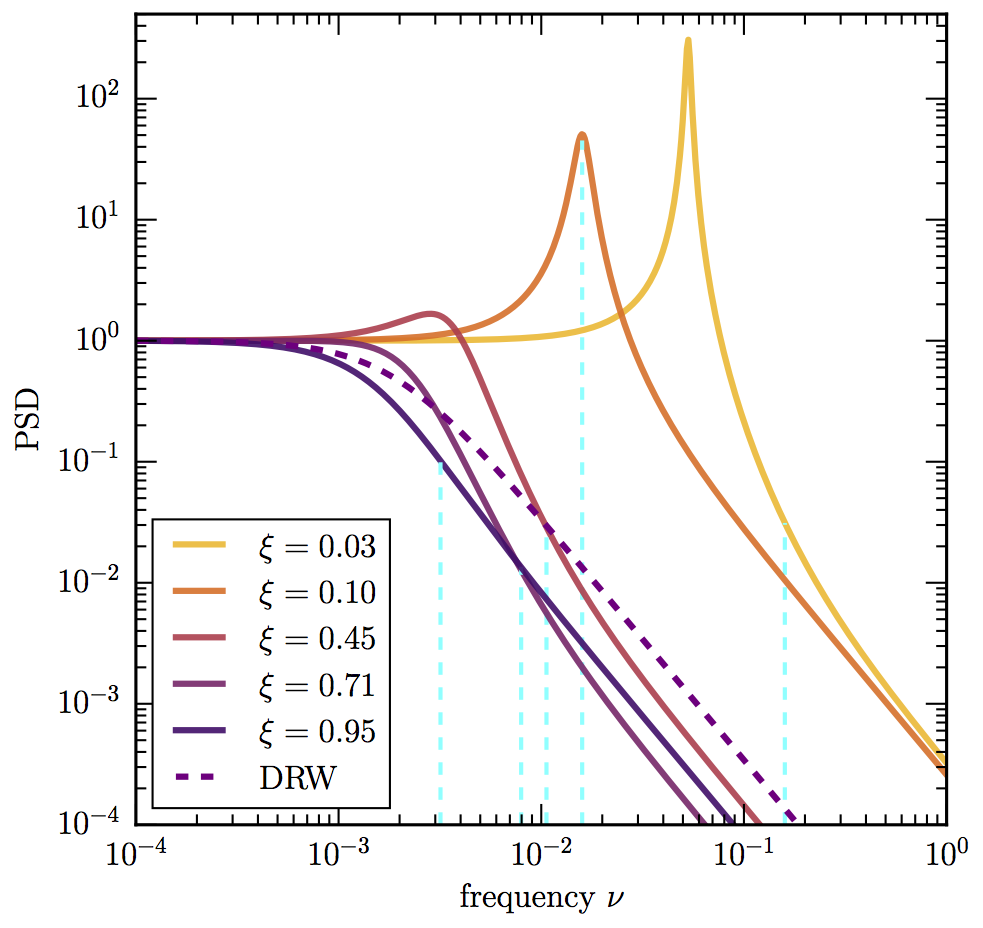}
    \caption{DHO PSDs with different characteristic frequencies and timescales. The left panel shows examples of DHO PSDs with real roots, while the right illustrates PSDs of models with complex roots.  Multi-sloped spectra can be modelled by the analytic PSD of DHO processes.  Dashed cyan lines track the blue noise perturbation frequency $\nu_{blue}$ which we increase for each curve yellow to red).
    The right panels show how a Lorentzian reveals the strength of a QPO signature. These examples demonstrate the flexibility of the DHO model, but they also provide a sense of how perturbation and response parameters interact to better recognize how the DHO model diverges from the DRW in the frequency domain.}
    \label{Fig:PSD}
\end{figure*}

The analytic PSD of a DRW (shown in the bottom panel of Figure~{\ref{Fig:PSD1}}) is parametrized by two values: (1) the asymptotic amplitude where the flat slope is observed, $\beta_{0}$, and (2) the decay timescale, $\tau$ in the C-AR parameter $\alpha_1$ (\citealt{KozlowskiDRW}, \citealt{MacLeod2008}, \citealt{Kelly14}), which are related by 
\begin{equation}
\label{eq:drwpsd}
PSD_{DRW}(\nu) = \frac{\beta_{0}^2}{\alpha_1 +(2\pi\nu^2)}.  
\end{equation}
The PSD of DHOs (CARMA(2,1)), on the other hand, involves two additional degrees of freedom, $\alpha_2$ and $\beta_1$,
\begin{equation}
\label{eq:dhopsd}
PSD_{DHO}(\nu) = \frac{1}{2\pi}\frac{\beta_{0}^2+4\pi^2 \beta_1^2 \nu^2}{16\pi^4\nu^4 +4\pi^2\nu^2 (\alpha_{1}^2- 2\alpha_2)+\alpha^2_2}    
\end{equation}
The numerator of Equation~\ref{eq:dhopsd} is the perturbation spectrum and therefore contains positively sloped components that add power, where as the denominator contains negatively sloped components since $\alpha_{1}$ and $\alpha_{2}$ relax and dissipate energetic perturbations to maintain equilibrium.

DRWs are scale invariant up to the decay timescale i.e. their PSDs follow single sloped power law across all observed frequencies in the lightcurve up to a break frequency where the spectrum flattens, $\nu_{flat}$.  A single-slope power law assumes self-similarity up to this frequency.  Self-similarity is broken when more complexity or covariance is present in the stochastic process. A secondary critical frequency may lead into a secondary slope in an empirical PSD. In the analytic PSDs of higher order CARMA processes, break frequencies or characteristic timescales can be estimated systematically by solving for critical points in the numerator and denominator of Eq.~\ref{eq:dhopsd}.  We show multi-sloped PSDs of DHOs in Figure~{\ref{Fig:PSD}}, however they are extensively flexible, thus this is not a complete preview.

The slopes of DHO PSDs can range from as steep as $Slope = -4$ to values with more power than the DRW, $Slope >-2$. They can also exhibit an additional characteristic frequency break to $\nu_{flat}$, which we will refer to as $\nu_{blue}$. The secondary frequency $\nu_{blue}$ reveals an 
excess power contribution such that frequencies higher than $\nu_{blue}$ have more power than the projected scaling relationship of lower frequencies. $\nu_{blue}$ is a critical point found by solving for the root of the numerator of Eq.~\ref{eq:dhopsd}, thus
$$\nu_{blue} = \frac{\beta_0}{2\pi\beta_1}.$$
Higher resolution optical data from Kepler (\citealt{Kepler}), Catalina Real Time Survey (CRTS; \citealt{Drake2009}), OGLE (\citealt{ogle}) and Pan-STARRS (\citealt{Kaiser2002}) show evidence of deviations from the broken power law in the form of excess power at high frequencies thus DHO models may prove useful to systematically measure the frequency where this deviation occurs (\citealt{Mushotzky2011}, \citealt{Zu2013}, \citealt{MGwavelet}, \citealt{Vish2015}, \citealt{Simm2016}).
The location of $v_{blue}$ for each example in Fig~\ref{Fig:PSD} is tracked by the dashed cyan lines until each terminates at the intersection with its corresponding model. 

The domain of PSD behavior is probably the most flexible in the case where the DHO fit results in two real roots $r_1$ and $r_2$. Two real roots are the negative inverse of the rise and decay timescales, $\tau_{rise}$ and $\tau_{decay}$.  When we examined Green's functions we learned that a rise timescale acts as a delay time over which perturbations drive the flux away from the mean.  
However, the Green's function only tells half of the story. In the left panel of Figure ~\ref{Fig:PSD}, we show how the rise mechanism couples to the perturbation frequency $\nu_{blue}$.   A dashed purple line shows an example DRW with a $\tau_{decay} = 150$ days. All solid curves (DHOs) have the same decay timescale and the same rise timescale, $\tau_{rise} = 25$. The only variable between each curve is the perturbation frequency $\nu_{blue}$.

When $\nu_{blue}$ is approximately equal to the magnitude of one of the real roots, then the rise timescale mechanism is cancelled out and we approximate the DRW model as shown by the yellow curve in the left panel. 
When $|r_1| < \nu_{blue} < |r_2|$, we obtain a more gradual turnover than the DRW, which is a more adequate model for AGN that show a slow flattening but no clear turnover.
If the perturbation frequency is higher than the magnitude of both roots, $\nu_{blue} > |r_1| \& |r_2|$, then the PSD is steeper than a DRW for low frequencies, until $\nu_{blue}$ where there is excess power than this scaling at the highest frequencies (shown by the red curve  in the left panel).

In the right panel of Fig~\ref{Fig:PSD}, we preview DHOs with complex roots with different damping ratios in the range $.01 <\xi < 1$.  When the 
the damping factor is $\xi <0.5$, the DHO PSD exhibits a Lorentzian as shown by the three lightest colored curves in the right panel.  The peaked-ness of the Lorentzian reveals how regular the periodicity is. The more rounded it appears, the less apparent the QPO will be in the lightcurve (hence ``quasi") (as seen previously in Fig~\ref{Fig:DampingFactor}).  If the $\nu_{blue}$ dominates frequencies lower than the QPO frequency, it will make the QPO resonate (more sharply peaked), especially if the damping factor is below the critical value. We can expect this combination of parameters when applying the DHO model to periodic stars (stars will have very low perturbation frequencies compared to quasars). X-ray binaries or pulsating stars with two oscillatory frequencies should employ fourth order CARMA models, which resolve two periodicities. 
QPO signatures in AGN although uncommon have been detected (\citealt{MGsmbh2015}, \citealt{Krista2018}). The DHO's ability to measure periodicities accurately should be further investigated.
We illustrate damped QPO features here as it is important to understand what the PSDs of potential AGN contaminants will look like if utilizing CARMA(2,1) not just for investigation of AGN physics, but also for AGN selection.

\section{Structure Functions }
\begin{figure} 
    \centering
    \label{sec:SF0}
	\plotone{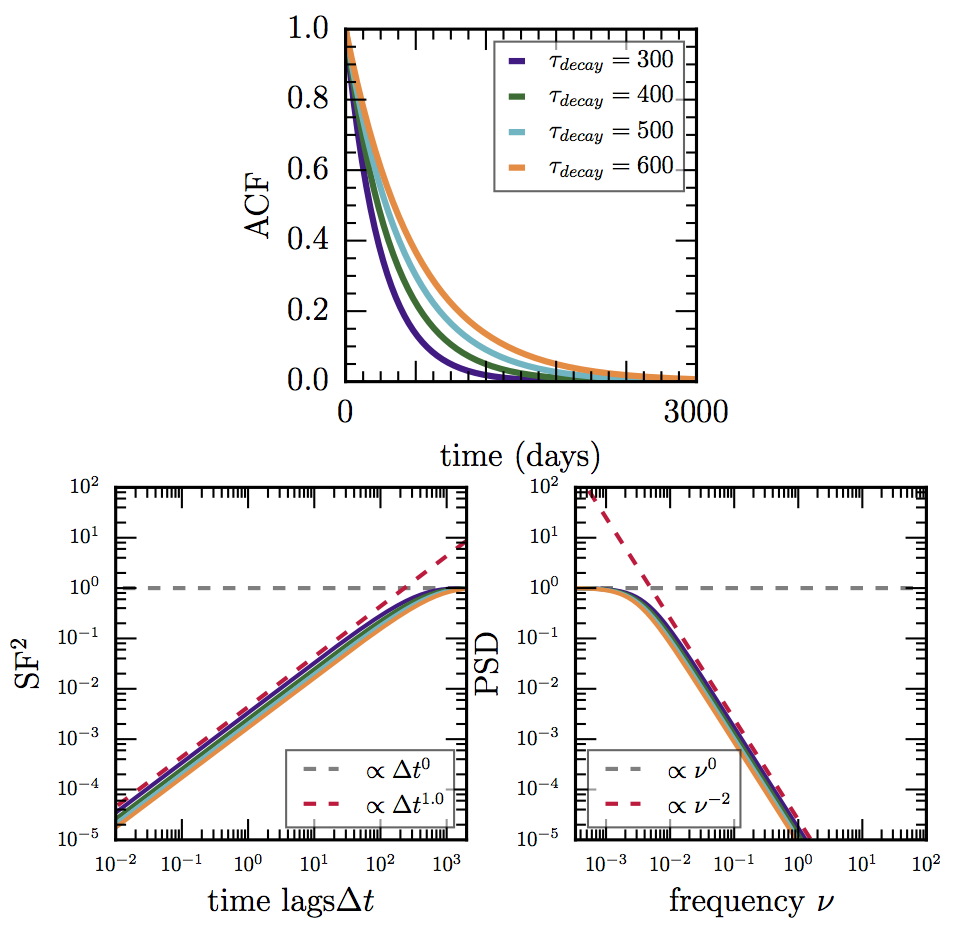}
    \caption{The ACF, SF,and PSD for various decay timescales. All panels reflect the same temporal baseline of 3000 days.  The projections of the SF and PSD are distorted mirror images of each other. Thus the decorrelation timescale ($\tau_{decorr}  \approx \frac{\pi}{2}\tau_{decay}$ of the DRW) is harder to observe in the SF and PSD for long timescales although the ACF fully decays within the same baseline.}
        \label{Fig:SF0}
\end{figure}

\begin{figure} 
    \label{sec:SF}
    \centering
	\includegraphics[width=.7\columnwidth]{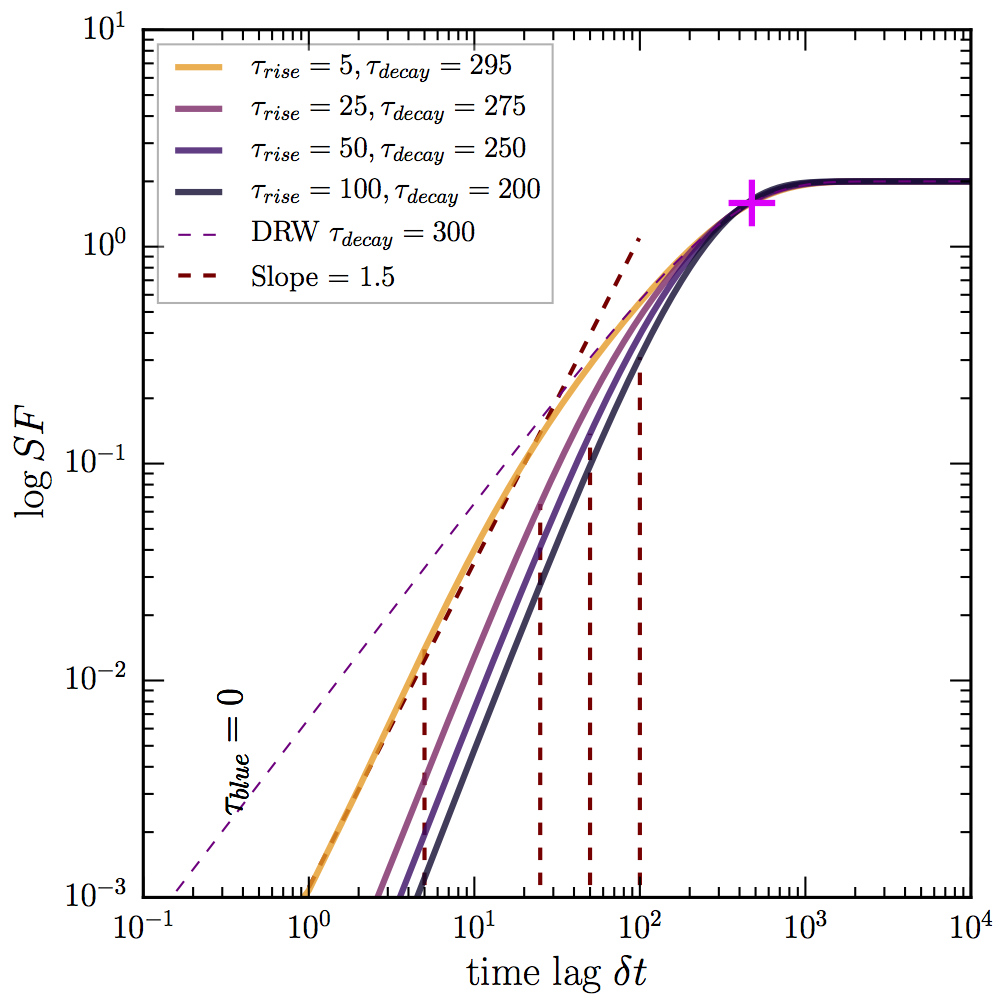}
	\includegraphics[width=.7\columnwidth]{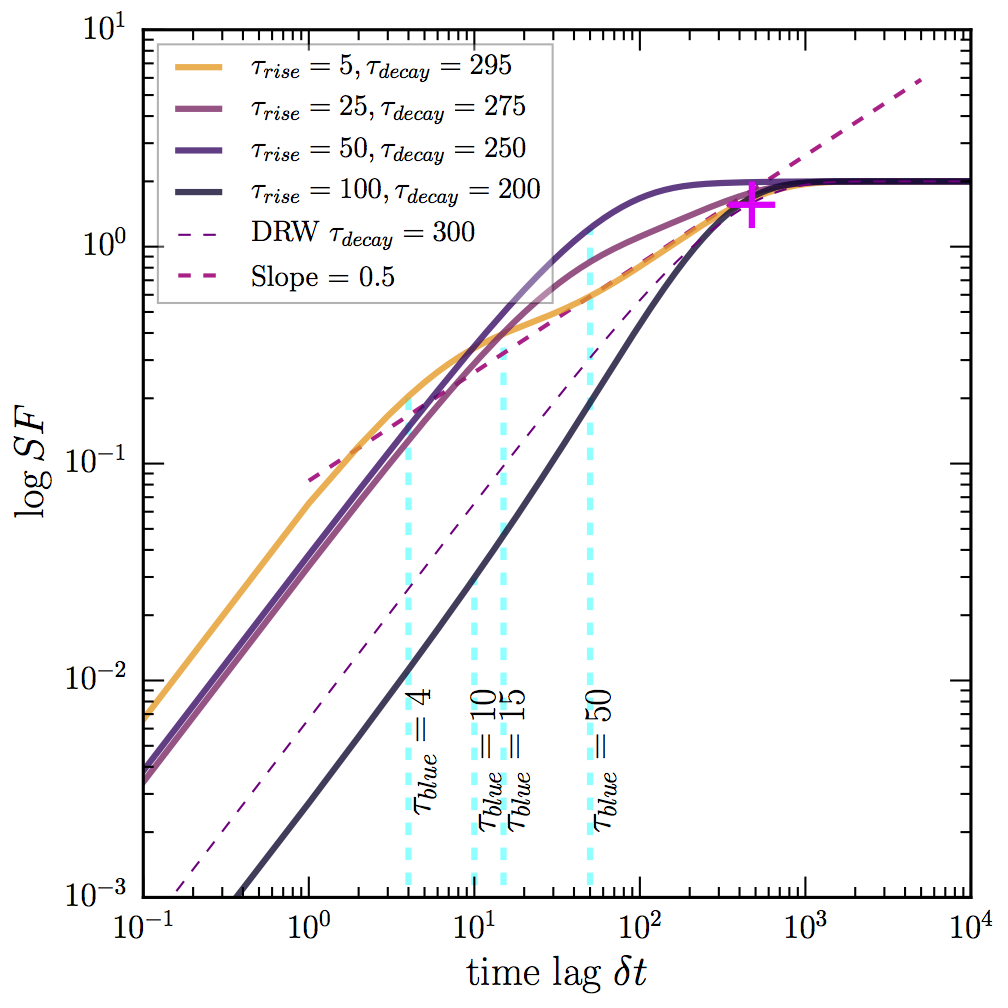}
	\includegraphics[width=.7\columnwidth]{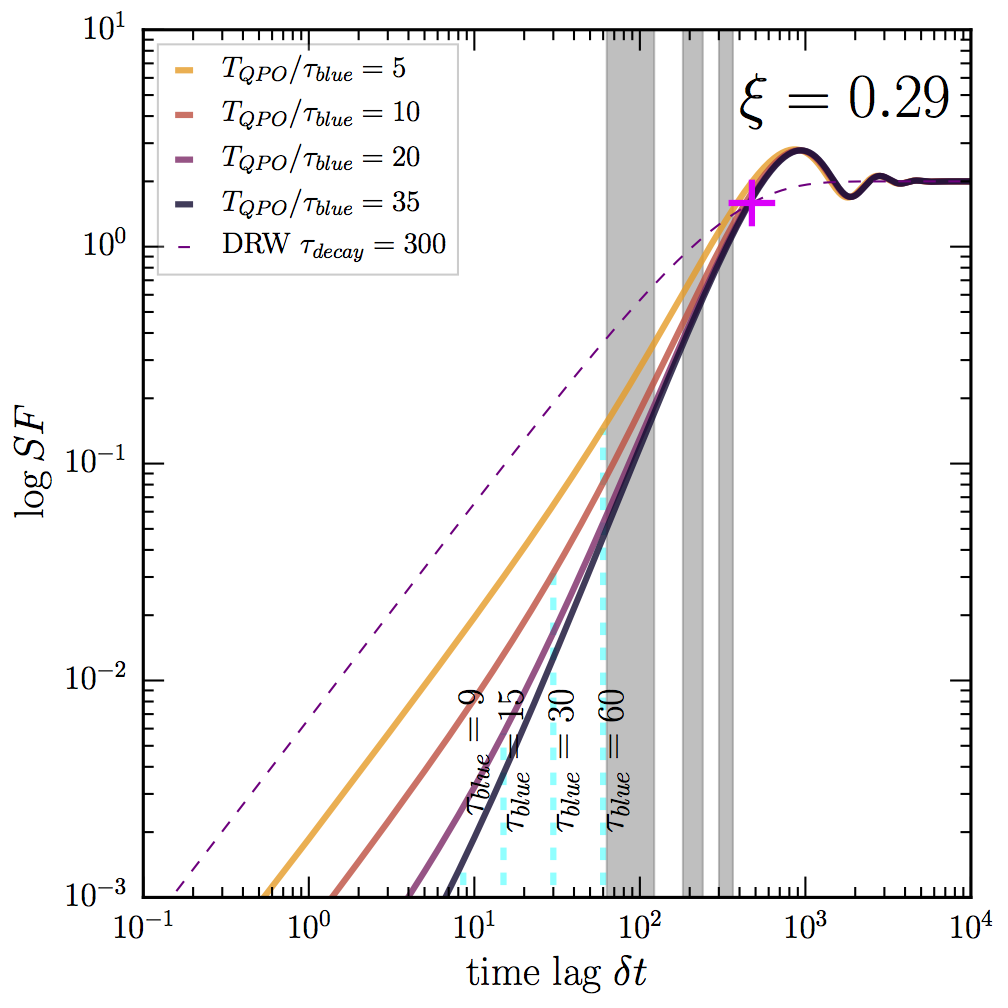}
    \caption{DHO SFs with different slopes. In the case of real roots (top two panels), the sum of the rise and decay timescale is related to the decorrelation timescale, but in the case of complex roots (bottom panel), the QPO timescale predicts the decorrelation timescale. DHO SFs exhibit slopes between 0 and 1.5. Generally real roots can produce SF with slope $<1$ and complex roots can predict slopes steeper than the DRW (slopes > 1 ) with possible excess at the timescale $\tau_{blue}$.  The grey fill (bottom panel) shows (6 month) observability gaps in the restframe of a quasar with redshift $z = 2$}
    \label{Fig:SF}
\end{figure}

The structure function (SF) is another common tool for the analysis of light curves; it is especially popular because it is more efficient to compute than a PSD for gappy astronomical data. A SF estimates the mean flux as a function of binned time lags of increasing length as depicted in Figure~\ref{Fig:SF0}.  The PSD and SF are both transforms of the ACF and appear like distorted mirror images of each other.  Since log-log projections stretch smaller values and concentrate together large orders of magnitude, the SF emphasizes the scaling of variability amplitude at relatively short timescales, while the PSD emphasizes correlation structure at low frequencies (or equivalently long timescales). As the ACF decay timescale of a stochastic lightcurve gets longer, then the turnover signaling decorrelation in PSDs is harder to resolve and estimate in SF space.     
Several definitions of the structure function are employed in the literature as well as critiques of the interpretation of SFs (\citealt{EmmSF2010}, \citealt{MacLeod2011}, \citealt{KozlowskiSF}). 
Here we review equivalent forms described by \citet{Vish2015} and \citet{KozlowskiSF} (first provided by \citet{SimonettiSF1985}). In parallel to the previous section on multi-sloped PSDs, we provide graphs of analytic DHO SFs in Figure~\ref{Fig:SF}.

The SF can be estimated directly from the data using the sample ACVF in Equation~\ref{eq:ACVF} and taking into account measurement error, $\sigma_{noise}$ as proposed by \citet{KozlowskiSF}
\begin{equation}
\label{eq:SF1}
SF_{obs}(\Delta t) = \sqrt{2(ACVF(0) - ACVF(\Delta t))-2\sigma_{noise}^2}.    
\end{equation}
The analytic SFs DRWs and DHOs is evaluated with nearly the same formula; however, the analytic ACVF described in Equations \ref{eq:acfdrw} and \ref{eq:acfdho} must be employed to compute each respectively,
\begin{equation}
\label{eq:SF2}
SF_{CARMA}(\Delta t) = \sqrt{2(ACVF(0) - ACVF(\Delta t))}    
\end{equation}
or equivalently,
\begin{equation}
\label{eq:SF3}
SF_{CARMA}(\Delta t) = \sqrt{2\sigma^2_{signal}(1-ACF(\Delta t))}.
\end{equation}
Uncertainty in SF$_{CARMA}$ can be accounted for by plotting 1-$\sigma$ confidence intervals of C-AR and C-MA parameters and a measurement noise floor at some amplitude cutoff (similar to PSD examples in \citealt{Kelly14} and \citealt{Vish2017}).  

Structure functions emphasize the information in the auto-correlation function at short lags, however, the highest priority timescale (as evidenced by the literature) of SF analysis is the decorrelation timescale of AGN (residing at long time lags). Another high priority statistic is the slope of a power law fitted to an empirical SF. A turnover to a flat slope in the SF (even if it is gradual) shows decorrelation just as in PSDs.  If there is no visible decay in the ACF, there is \textit{no chance} of recovering any characteristic timescale in the SF because the process is non-stationary. However, if there is a decay in the ACF, then timescale estimation is possible even if there is no clear turnover in the empirical SF or PSD. These timescales will not indicate decorrelation but rather they reveal either deviations from the DRW SF or entirely different slopes ranging from $0 < Slope < 1.5$.  

The decorrelation timescale is related to the DRW decay timescale by the relation $\tau_{decorr} \propto c\tau_{decay}$, where $c$ is a numerical constant. A SF (or PSD) exhibiting a more gradual turnover than a DRW SF at long lags may be better described by a multi-sloped model. If we do not assume a broken power law, but rather a multi-sloped SF (or PSD) with a slow flattening towards an asymptotic amplitude, we would seek more than one characteristic timescale rather than one signature turnover.

Analytic structure functions of DHOs with multi-sloped power-laws are illustrated in Figure~\ref{Fig:SF}. DHOs with real roots which provide a rise timescale $\tau_{rise}$ in addition to a decay timescale $\tau_{decay}$ are displayed in the top two panels and DHOs with complex roots which provide a QPO period $T_{QPO}$ (rather than a rise timescale) are shown in the bottom panel. 

The pink cross hairs on each panel of Figure~\ref{Fig:SF} are located at the time lag where the SF amplitude is equal to $0.795 SF_{\infty}$, which was empirically shown by \citet{KozlowskiSF} to be a model-independent method to estimate the decorrelation timescale  
$$\tau_{decor} = \Delta t(0.795 SF_{\infty})$$
of SFs. If a DHO fit to a lightcurve results in two real roots, then the decorrelation timescale of the SF can be estimated as
$$\tau_{decorr} \approx \frac{\pi}{2}(\tau_{rise}+\tau_{decay}).$$
This relation is illustrated in the first panel which shows that various sums of $\tau_{rise}$ and $\tau_{decay}$ totaling to $300$ days all intersect at the same turnover timescale (marked by the pink cross-hairs) including a DRW walk with a decay timescale of 300 days. When the DHO fit results in complex roots the decorrelation timescale is computed differently, 
$$\tau_{decorr} \approx \frac{\pi}{2}T_{QPO}.$$ 
We provide these relations to reconcile the timescales of the DHO to timescales of SF analyses that are independent of CARMA modelling \footnote{These relations are only valid when $\tau_{blue} < (\tau_{rise} \text{ or } T_{QPO}$)}.

The flexibility of the DHO SF is very extensive; we choose to highlight its ability to model different slopes. The DRW slope is fixed, but DHO processes are described by a range of slopes $0 \rightarrow 1.5$.  When the perturbation timescale is very close to zero, both $\tau_{rise}$ and $T_{QPO}$ mark critical points where the slope is steeper than the DRW.  
When $\tau_{blue}$ becomes significant relative to the value of $\tau_{rise}$, then timescales greater than $\tau_{blue}$ exhibit slopes $ < 1$  (diagrammed in the middle panel). The effect is opposite in the complex case,  when $\tau_{blue}$ becomes significant relative to the value of $T_{QPO}$, slopes are steeper ($>1$) than the DRW. If $\tau_{blue} \geq (\tau_{rise}$ \text{ or } $T_{QPO}$) this condition is a test of overfitting and the DRW may be more adequate to describe the lightcurve.

\section{Discussion and Summary}
\begin{figure}
    \centering
    \plotone{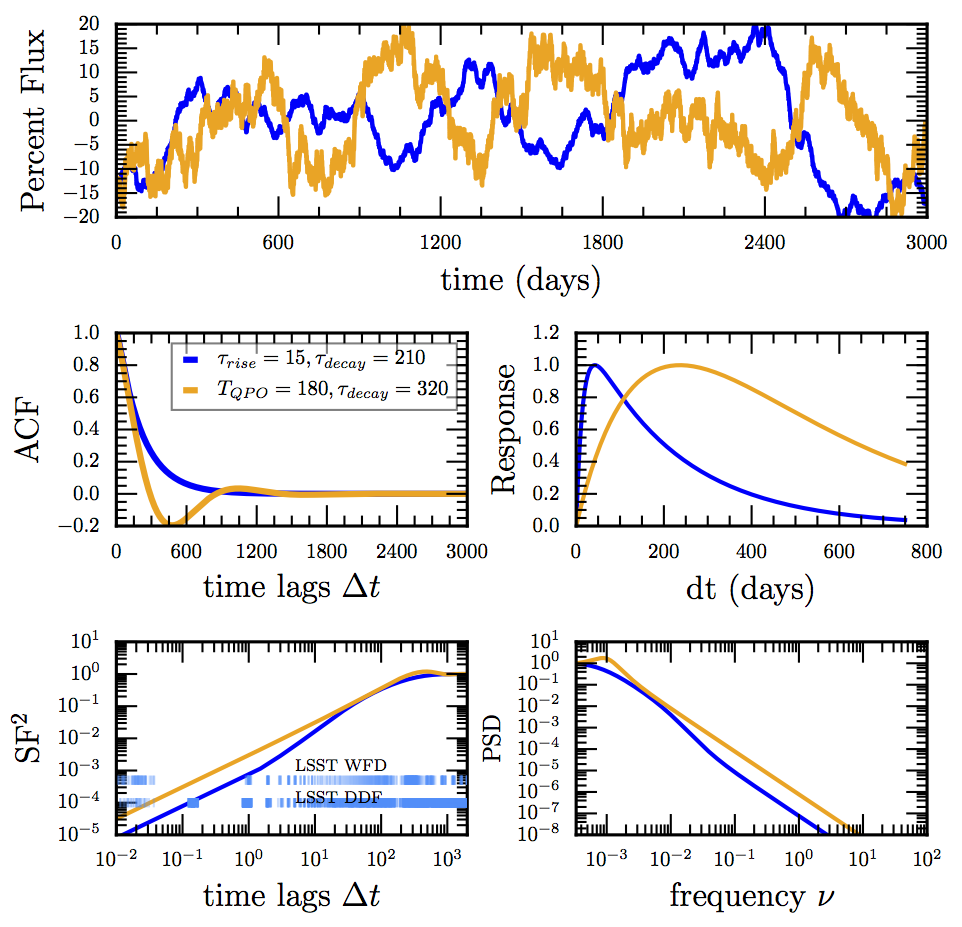}
    \caption{AGN variability analysis toolset: CARMA models, ACFs, Impulse-Response functions, SFs and PSDs. We show a sample DHO model with complex roots in orange and a sample DHO with real roots in blue.  Both examples diverge the most at short time lags in the ACF and SF.  The low frequency discrepancies although subtle are magnified in the response function and PSD. Direct lightcurve modelling provides parameters to compute analytic SFs and PSDs, that correspond to flexible (multi-sloped) power models, which can be compared with empirical estimates of the SF and PSD.}
    \label{fig:toolkit}
\end{figure}

This work provides a guide for the application and interpretation of auto-regressive (CARMA) models studied jointly with causal Green's Functions for the investigation of AGN variability.  Directly modelling a lightcurve with a stochastic process of flexible complexity extracts more descriptive information at short and long timescales. To better understand what C-AR and C-MA parameters are capable of describing, we investigated the second order statistics of DRWs and DHOs including the ACVF, PSD, and SF---which are used extensively to study astrophysical variability (see AGN variability analysis toolset in Figure~\ref{fig:toolkit}).  The simplest model of stochastic variability is a power law in the SF or PSD that decorrelates at long timescales. 
A deviation from a single-slope power law in an AGN SF or PSD indicates higher covariance which may be well described by a CARMA process. 

C-AR parameters describe a dynamic system (by defining a differential equation) and C-MA parameters perturb the system. In other words, we can generally predict the motion of a physical system via two delay timescales (motion away or toward equilibrium).  However, perturbations make the motion more random and less predictable depending on the amplitude and frequency of the perturbation mechanism. These delay timescales as well as the perturbation frequency (or timescale) come from the roots of the CARMA differential equation. They can be interpreted with the Green's function, ACF, PSD and SF as we have shown, and they are related to critical points in the analytic formulas where the PSD and SF deviate from the single-slope model, i.e., stochastic processes with more characteristic timescales produce multi-slope power laws.

We discussed the multi-sloped power-law model applied to PSDs and SFs as a method of conducting multi-scale analysis.  Multi-scale analysis may disentangle the significance of multiple characteristic timescales or frequencies found in observations of the same AGN that have different time resolution (cadence).  Multiple timescales may indicate several processes dominating at different length scales or a break from homogeneity in a control parameter, such as density or temperature, from one length scale to the next.  We also propose employing techniques used to study critical phenomena to interpret the (ARMA) DHO triangle and to approach ensemble studies of AGN in the CARMA parameter landscape.  

In ensemble studies, different positions in the CARMA parameter space may be related to different mechanisms involved in driving variability. It merits further investigation to understand why some roots (or solutions) may enter the complex plane while others reside only in the real regime. Dynamic systems with complex roots may have high dimensionality ($N > 2$) in the number of control parameters driving variability or they may be a systematic of non-stationarity. See the appendix for a summary of CARMA DHO timescales.

The next steps towards improving AGN variability analyses techniques are to carefully investigate the accuracy of timescale estimation from gappy lightcurves. Tests are needed to identify systematic biases in timescale estimation depending on cadence and length, following the example of \citet{KozlowskiLength} with the DRW timescale.  We will further investigate degeneracies between the DRW and DHOs and high proximity models in the SF and PSD space using the analytic formulations provided by CARMA analyses.  Provided improvements in sampling resolution, measurement error and improved techniques for combining surveys (to obtain extended baselines), these advancements in stochastic modelling may provide meaningful timescales that probe the accretion physics of these diverse and extreme engines. 

\section{Acknowledgements}
We acknowledge support from NASA grant NNX17AF18G.
We wish to thank Vishal Kasliwal for his input, discussions and the development of the time series modelling software, Kali. We thank David Lioi and Mark Davis for theoretical discussions of applied methods.
We wish to thank Weixiang Yu for valuable input and discussions regarding LSST cadences. Thanks to the LSSTC Data Science Fellowship Program for educational support.

\bibliographystyle{./yahapj}
\bibliography{./ms}

\section{Appendix}
We summarize CARMA DHO timescales here: 
\begin{itemize}
\item Rise timescale (real roots)
$$\tau_{rise} = \frac{1}{r_{2}-r_{1}}ln |\frac{r_1}{r_2} | \approx |\frac{1}{min(r_1,r_2)}|$$

\item Effective Decay timescale of the lightcurve 
$$\tau_{effdecay} = \frac{G_{t_{rise}}}{e} \approx \tau_{rise}+\tau_{decay},$$  

$$\tau_{decay} = |\frac{1}{max(r_1,r_2)}|$$

\item QPO natural period  (complex roots) 
$$T_{QPO} = |\frac{1}{\Im(r_1)}|= |\frac{1}{\omega_0}| \approx \frac{2\pi}{\sqrt{\alpha_{1}}}$$

\item QPO damped period  (complex roots) 
$$T_{dQPO} = |\frac{2\pi}{\omega_d}| = \frac{2\pi}{\omega_0\sqrt{1-\xi^2}}$$

with resonance depending on the damping ratio,
$$\xi = \frac{\alpha_1}{2\sqrt{\alpha_2}}$$

\item Coherence/Decay timescale (complex roots) describes how resonant the QPO is
$$\tau_{decay}= |\frac{1}{\Re(r_1)}| = |\frac{1}{\Re(r_2)}|$$

\item Blue noise dominates at the timescale,
$$\tau_{blue} = |\frac{\beta_1}{\beta_0}|$$

\item Multi-slope SF decorrelation timescale
$$\tau_{decorr} \approx  \frac{\pi}{2}(\tau_{rise}+\tau_{decay}) \text{ (or) } \frac{\pi}{2}T_{QPO}$$

\item SF Secondary slope timescale
$$\tau_{excess} = \frac{\tau_{blue}}{\sqrt{2}\pi} $$ 
or
$$\tau_{steep} = \tau_{rise} \text{ when }\tau_{blue} << \tau_{rise} $$

\item PSD secondary slope 
$$\nu_{blue} = \frac{\beta_1}{2\pi\beta_0}$$
$$\text{ when }\tau_{blue} << \tau_{rise} \text{ or } T_{QPO}$$
$$PSD \sim \nu^{-2} \rightarrow \nu^{-4}$$
$$\nu_{steep} = min(\frac{2\pi}{\sqrt{\alpha_1^{2}-2\alpha_2}}, \frac{2\pi}{\sqrt{\alpha_2}})$$
\end{itemize}

\end{document}